\documentclass{aa}  
\usepackage{multicol}
\usepackage{graphicx}
\usepackage{placeins}
\usepackage{lscape}
\usepackage{txfonts}
\usepackage{natbib}
\usepackage{amsmath}
\usepackage[usenames]{color}
\usepackage[dvipsnames]{xcolor}
\usepackage{comment}
\usepackage{hyperref}
\hypersetup{
    colorlinks=true,
    citecolor =blue,
    linkcolor=blue,
    filecolor=black,      
    urlcolor=blue} 

\begin{document} 

   \title{Multi-band infrared imaging reveals dusty spiral arcs around the binary B[e] star \object{$3$~Puppis}}

   \author{M.~Abello\inst{1}\fnmsep\thanks{Corresponding author: margaux.abello@oca.eu}, J.~Drevon\inst{2}, A.~Meilland\inst{1}, A.~Domiciano de Souza\inst{1}, F.~Millour \inst{1}, R.~Flor\inst{3}, J. H.~Leftley\inst{4}, C.~Paladini\inst{2}, Ph.~Stee\inst{1}, A.~Matter\inst{1}, S.~Lagarde\inst{1}, B.~Lopez\inst{1}, P.~\'Abrah\'am\inst{5,6,7}, J.-C.~Augereau\inst{8}, P.~Cruzalèbes\inst{1}, W.~Danchi\inst{9}, T.~Henning\inst{10}, T.~Juhász\inst{6}, F.~Kerschbaum\inst{5}, F.~Lykou\inst{6}, P.~Priolet\inst{8}, S.~Robbe-Dubois\inst{1}, J.~Varga\inst{6}, L.B.F.M.~Waters\inst{11,12},  G.~Weigelt\inst{13}, S.~Wolf\inst{14} 
					\and
					MATISSE collaboration}
          
    \institute{Université Côte d'Azur, Observatoire de la Côte d'Azur, CNRS, Laboratoire Lagrange, Boulevard de l'Observatoire, CS 34229, 06304 Nice cedex 4, France
        \and
             European Southern Observatory, Alonso de Córdova, 3107 Vitacura, Santiago, Chile
        \and
            ONERA / DEMR, Université de Toulouse, F-31055 Toulouse
        \and 
            Department of Physics \& Astronomy, University of Southampton, Southampton, SO17 1BJ, UK
        \and
            Institute for Astronomy (IfA), University of Vienna, T\"urkenschanzstrasse 17, A-1180 Vienna, Austria
        \and 
            HUN-REN Research Centre for Astronomy and Earth Sciences, Konkoly Observatory, MTA Centre of Excellence, Konkoly Thege Miklós út 15-17., H-1121 Budapest, Hungary
        \and
            Institute of Physics and Astronomy, ELTE E\"otv\"os Lor\'and University, P\'azm\'any P\'eter s\'et\'any 1/A, 1117 Budapest, Hungary
        \and
            Université Grenoble Alpes, CNRS, IPAG, 38000 Grenoble, France
        \and
            NASA Goddard Space Flight Center, Astrophysics Division, Greenbelt, MD, 20771, USA
        \and 
            Max-Planck-Institut für Astronomie, Königstuhl 17, 69117 Heidelberg, Germany
        \and
            SRON Netherlands Institute for Space Research, Niels Bohrweg 4, 2333 CA Leiden, The Netherlands
        \and 
            Center for Astrophysics Harvard \& Smithsonian, 60 Garden Street, Cambridge, MA 02138, USA
        \and 
            Max Planck Institute for Radio Astronomy, Auf dem Hügel 69, D-53121, Bonn, Germany
        \and 
            Institute of Theoretical Physics and Astrophysics, University of Kiel, Leibnizstraße 15, 24118, Kiel, Germany
    }

   \date{Received 26 July 2025 / Accepted 24 October 2025}
   \titlerunning {Multi-band infrared imaging reveals dusty spiral arcs around the binary B[e] star \object{$3$~Puppis}}
   \authorrunning {Abello et al.}
 
  \abstract
  {\object{$3$~Puppis} stands out as the brightest star known exhibiting the B[e] phenomenon. Although recent studies have classified this A-type star within the supergiant group, the influence of its binary nature on the circumstellar environment (CE) remains difficult to model and interpret.}
  {To resolve the dusty regions of \object{$3$~Puppis} at angular scales of 5--10\,milliarcseconds (mas), we conducted high-angular-resolution interferometric observations with the mid-infrared beam combiner VLTI/MATISSE across the $3$--$12~\mu$m wavelength range.}
  {Since the $(u,v)$ coverage is sufficient to perform image reconstruction, we present an innovative statistical interferometric imaging technique based on the \texttt{MiRA} software to produce averaged images. Applied to MATISSE data, this systematic approach enables the selection of an optimal set of reconstructions, improving the robustness and fidelity of the recovered features. We also apply \texttt{SPARCO}, an independent imaging tool well suited to systems with a bright central source surrounded by a fainter and extended CE.}
  {The images obtained with both tools in the L, M, and N spectral bands show good agreement and clearly reveal an asymmetric elongated structure located at $\sim17$~mas ($\sim10$~au at 631~pc) to the southeast of the image centre, with a density contrast of $\sim20\%$. A second asymmetry to the northwest and a skewed inner rim are also detected. Simple geometric modelling, inspired by the reconstructed images, provides quantitative constraints on the morphology, position, and flux contribution of the CE and its asymmetries.}
  {Our final MATISSE images are consistent with previous VLTI interferometric observations while revealing a more complex CE with large-scale clumps in the southeast and northwest disc regions of \object{$3$~Puppis}. Finally the hydrodynamic simulation concludes that the tidal spiral wake perturbations driven by the central binary, dynamically excited at Lindblad resonances within the circumbinary disc, provide the best interpretation for the radial extent and curvature of the elongated structures observed across all bands.} 
  
   \keywords{techniques: high angular resolution -- techniques: interferometry -- stars: emission-line, Be -- circumstellar matter --}

   \maketitle

\section{Introduction}\label{sec:intro}
An enigmatic and highly debated group of stars surrounded by a rich circumstellar environment (CE) are the so-called B[e] stars --properly defined as emission-type stars exhibiting the B[e] phenomenon. First disentangled from classical Be stars by \cite{Allen_Swings1976}, B[e] stars typically have spectral types ranging from O9 to A2, and are embedded in compact ionized gaseous discs and surrounded at larger scales by extended dusty envelopes. So far, 120 candidates across the Galactic plane and Magellanic Clouds have been identified according to the latest report of \cite{Chen+2021}. However, many B[e] stars remain poorly studied. Therefore, their evolutionary status, disc geometry, CE structure, kinematics, and underlying physical processes remain uncertain or sometimes not even understood for some of them.\\
Despite sharing similar physical CE properties (e.g. temperature, matter density), they represent a heterogeneous classification. Indeed, they encompass objects across various evolutionary stages, from pre- to post-main sequence stars (e.g. Herbig Ae/Be, compact planetary nebulae), and spanning a wide mass range, from sub-solar masses up to 70 $M_\odot$ \citep{Liimets+2022}, including B[e] supergiants (SG, \citealp{Zickgraf+1986}).\\
The spectral features of the B[e]-star population are characterized by the simultaneous presence of low-excitation forbidden and permitted emission lines (e.g. [OI], [CaII], H$_\alpha$, Br$_\gamma$) in their spectra \citep{Lamers+1998, Miroshnichenko_2007}. Molecules can also be present, in particular in the region between the atomic gas and the dust (e.g. CO, SiO, TiO). Therefore, the rich chemistry of their CE results in a strong infrared (IR) excess.\\
Two main scenarios are generally invoked to explain the origin of this CE: (a) decretion mechanisms driven by the star itself (e.g. the bi-stability model proposed by \citealp{Lamers_Pauldrach1991}) and (b) mass-transfer episodes triggered by a close companion (proposed by \citealp{Miroshnichenko_2007} and discussed more broadly in \citealp{Ellerbroek+2015}). Other processes might also contribute in some specific cases, for instance pulsation-driven mass loss \citep{Baade+2016} or perturbations induced by a third body \citep{Michaely_Perets2014}. \\
Since a number of binaries have been discovered among the unclassified B[e] stars over the years (evaluated to $\sim 30\%$ in \citealp{Varga+2019}), \cite{Miroshnichenko_2007} has established a new category of B[e] stars that gathers the binary systems for which the CE has been created by an earlier mass transfer: the \object{FS~CMa} group. Binary interaction provides a possible explanation for the observed high mass-loss rates \citep{Kraus2016, Kraus+2016} and is therefore seriously debated in the community. However, the structure of a CE produced by a binary component is even more complex to model and understand. These objects therefore greatly benefit from high-angular-resolution observations. Indeed, to spatially resolve the innermost regions of B[e] stars --thus providing a direct measurement of sizes, positions, and relative fluxes-- we need facilities able to see the milliarcsecond (mas) scale, so far only accessible through long-baseline optical (i.e. visible and IR) interferometry. In this work, we use the latest-generation mid-IR interferometric beam-combiner at the Very Large Telescope Interferometer (VLTI) to spatially resolve the intricate CE structure of \object{$3$~Puppis} (a.k.a. \object{$3$~Pup}, \object{$l$~Pup}, \object{HD~62623}), the binary in our Galaxy that stands out as the brightest known object exhibiting the B[e] phenomenon.\\
Based on radial velocity variations, \cite{Rovero_Ringuelet1994} and \cite{Plets+1995} proposed that \object{$3$~Pup} could be an interacting binary surrounded by discs made of gas and dust. \citet{Miroshnichenko+2020} supported this view and suggested that \object{$3$~Pup} is an FS~CMa star, composed of an intermediate mass A[e] SG and a low-mass companion, both embedded in a gaseous circumbinary disc. In this scenario, the complex gaseous and dusty CE would result from one or more mass transfer events within the binary system. Previous IR spectro-interferometric observations have placed constraints on the geometry and relative flux contributions of the gas and dust CE components, and have confirmed that the CE contains a disc-like structure in Keplerian rotation \citep{Meilland+2010,Millour+2011}. On the other hand, the distance to \object{$3$~Pup} remains difficult to constrain.  \\
Historically, the usually adopted value is around 600--700~pc, estimated through spectroscopic or BCD spectrophotometric studies (e.g. $630 \pm 85$~pc, \citealp{Chentsov+2010}) rather than the parallax inversion since in the pre-\emph{Gaia} period the latter yielded unreliable results (e.g. identified in \emph{Hipparcos} catalogue as \object{HIP~37677}, the parallax estimation produces a invalid negative distance due to uncertainties larger than the parallax measurement). While the second data release of the \emph{Gaia} astrometry provides a similar trigonometric distance estimation ($d_\mathrm{GeDR2}\simeq 631^{+159}_{-106}$~pc, \citealp{GaiaDR2_2018}), the third release gives a significantly different value ($d_\mathrm{GeDR3}\simeq 1\,028^{+206}_{-147}$~pc, \citealp{DR3Gaia_Babusiaux+2023}). \cite{Aidelman+2023} explain that this discrepancy is expected for B[e] stars because the extended CEs can shift the overall photocentre position detected by \emph{Gaia} astrometry, leading to deviations and degradation in the parallax measurements. This remark is consistent with the renormalised unit weight error returned in \cite{RUWE_GaiaEDR3_2023} (i.e. RUWE = $2.664$). Thus, a reliable distance cannot be derived by directly inverting the raw \emph{Gaia} parallax, even in DR3 with its improved calibration systematics. \\
The probabilistic Bayesian distance estimation method has been proposed to overcome this effect \citep{Bailer_Jones_EDR3_2021} by adding colour–magnitude information, yielding the photogeometric distance estimate generally with higher accuracy and precision for stars with poor parallaxes. Assuming an unimodal source (i.e. a single star in our Galaxy), the corresponding median value of the posterior probability distribution is $d_\mathrm{PGeDR3}\simeq 1\,024^{+231}_{-121}$~pc, with $68\%$ confidence interval around this value. Even if the Bailer-Jones photogeometric distance is likely the best available estimate from \emph{Gaia} data alone for \object{$3$~Pup}, it should still be used with caution since its unusual characteristics and nature may not be well-represented in the Galactic model priors, and the binary nature violates the single-star assumption.\\
Since the interferometry technique provides direct physical measurements, the choice of either distance will not affect our results. We adopt the \emph{Gaia} DR2 value for consistency with the \cite{Miroshnichenko+2020} study, but we also present results using the photogeometric distance for comparison. Table~\ref{tab:params_prop_lPup_litterature} summarizes the fundamental stellar parameters and the CE properties of \object{$3$~Pup} derived from previous studies. \\
The paper is structured as follows. The observations are presented in Sect.~\ref{sec:obs}, as well as the data reduction and calibration processes applied. The interferometric image reconstruction techniques adopted and the imaging results obtained for \object{$3$~Pup} are presented in Sect.~\ref{sec:method_imaging}. A discussion of these results and the conclusions of this work are respectively given in Sects.~\ref{sec:discuss} and \ref{sec:concl}.\\
\begin{table}[ht]
\caption{List of selected stellar parameters and circumstellar environment properties of the \object{$3$~Puppis} system from the literature.}
\begin{tabular}{llc}
\hline\hline 
\noalign{\smallskip}
Parameter & Value & Ref. \\
\noalign{\smallskip}
\hline
\noalign{\smallskip}
Distance $d_\mathrm{GeDR2}$ (pc)& $631^{+159}_{-106}$ &1 \\
\noalign{\smallskip}
Distance $d_\mathrm{PGeDR3}$ (pc)& $1\,024^{+231}_{-121}$ & 7\\
\noalign{\smallskip}
Luminosity $log(L_*/L_\sun)$ & $4.1 \pm 0.1$ &2\\
\noalign{\smallskip}
Relative flux $F$ & $F_* = 34 $, $F_\text{gas} = 11$ &3\\
($\%$ K-band total flux$^{**}$)& $F_\text{dust} = 35$, $F_\text{bkd} = 20$&3\\
\noalign{\smallskip}
\hline
\noalign{\smallskip}
\multicolumn{3}{c}{Supergiant star} \\
\noalign{\smallskip}
\hline
\noalign{\smallskip}
Spectral type & A2.7Ib &2\\
\noalign{\smallskip}
Radius $R_*$ ($R_\sun$) & $54 \pm 7$ &2\\
Radius $R_*$ (mas) & $0.40 \pm 0.07$&1, 2\\
\noalign{\smallskip}
Temperature $T_\mathrm{eff,*}$ (K) & $8500\pm500$&2\\
\noalign{\smallskip}
Mass $M_*$ ($M_\sun$)& $8.8 \pm 0.5$ &2\\
\noalign{\smallskip}
\hline
\noalign{\smallskip}
\multicolumn{3}{c}{Companion star} \\
\noalign{\smallskip}
\hline
\noalign{\smallskip}
Radius $R_\mathrm{c}$ ($R_\sun$)& $\sim 0.3$ &2\\
\noalign{\smallskip}
Temperature $T_\mathrm{eff,c}$ (K)& $\sim 50\,000$ &2\\
\noalign{\smallskip}
Mass $M_\mathrm{c}$ ($M_\sun$)& $0.75 \pm 0.25$ &2\\
\noalign{\smallskip}
Orbital period $P_\mathrm{orb}$ (days)& $137.52 \pm 0.04$ &6\\
\noalign{\smallskip}
Separation $a$ (au)& $1.11 \pm 0.03$ &2\\
Separation $a$ (mas) & $1.76 \pm 0.05$ &1,2\\
\noalign{\smallskip}
\hline
\noalign{\smallskip}
\multicolumn{3}{c}{Gaseous circumstellar envelope} \\
\noalign{\smallskip}
\hline
\noalign{\smallskip}
Extension $D_\text{gas}$ (mas)& $\sim 1.39$ &3\\
Extension $D_\text{gas}$ (au) & $0.877^{+0.221}_{-0.147}$ &1,3\\
\noalign{\smallskip}
Mass $M_\text{gas}$ ($M_\sun$) & $\sim 4 \times 10^{-7}$ &4\\
\noalign{\smallskip}
\hline
\noalign{\smallskip}
\multicolumn{3}{c}{Dusty circumstellar disc} \\
\noalign{\smallskip}
\hline
\noalign{\smallskip}
Inner rim radius $R_\mathrm{rim}$ (mas)& $ 5.7 \pm 1.2$ & 4,3\\ 
Inner rim radius $R_\mathrm{rim}$ (au)& $3.6^{+1.1}_{-0.9}$ & 1,4 \\
\noalign{\smallskip}
Mass $M_\text{dust}$ ($M_\sun$)& $\sim 3.5 \times 10^{-7}$ &4\\
\noalign{\smallskip}
Dust temperature $T_\mathrm{rim}$ (K)& $\sim 1\,250$ (SiO) &4\\
\noalign{\smallskip}
Major-axis angle $\mathrm{PA}$ ($\degr$)& $15 \pm 10$ &4\\
\noalign{\smallskip}
Inclination angle $i$ ($\degr$)& $60 \pm 10$ &4\\
\noalign{\smallskip}
Elongation ratio $e$ (...) & $1.3 \pm 0.1$ &4\\
\noalign{\smallskip}
Polarization angle $\theta_\mathrm{pol}$ ($\degr$)& $95 \pm 5$ &5\\
\noalign{\smallskip}
\hline
\end{tabular}\label{tab:params_prop_lPup_litterature}
\tablebib{(1)~derived from \emph{Gaia} DR2 parallax \citep{GaiaDR2_2018}; (2)~\citet{Miroshnichenko+2020}; (3)~\citet{Millour+2011}; (4)~\citet{Meilland+2010}; (5)~\citet{Yudin_Evans1998}; (6)~\citet{Vaidman+2025} ; (7)~ from the \citeauthor{Bailer_Jones_EDR3_2021} catalogue; $^{**} F_\mathrm{tot,K} = 69$ Jy from \citeauthor{MDFC_Cruzalebes+2019} catalogue.}
\end{table}

\begin{figure}[ht]
   \centering
   \includegraphics[width=\linewidth]{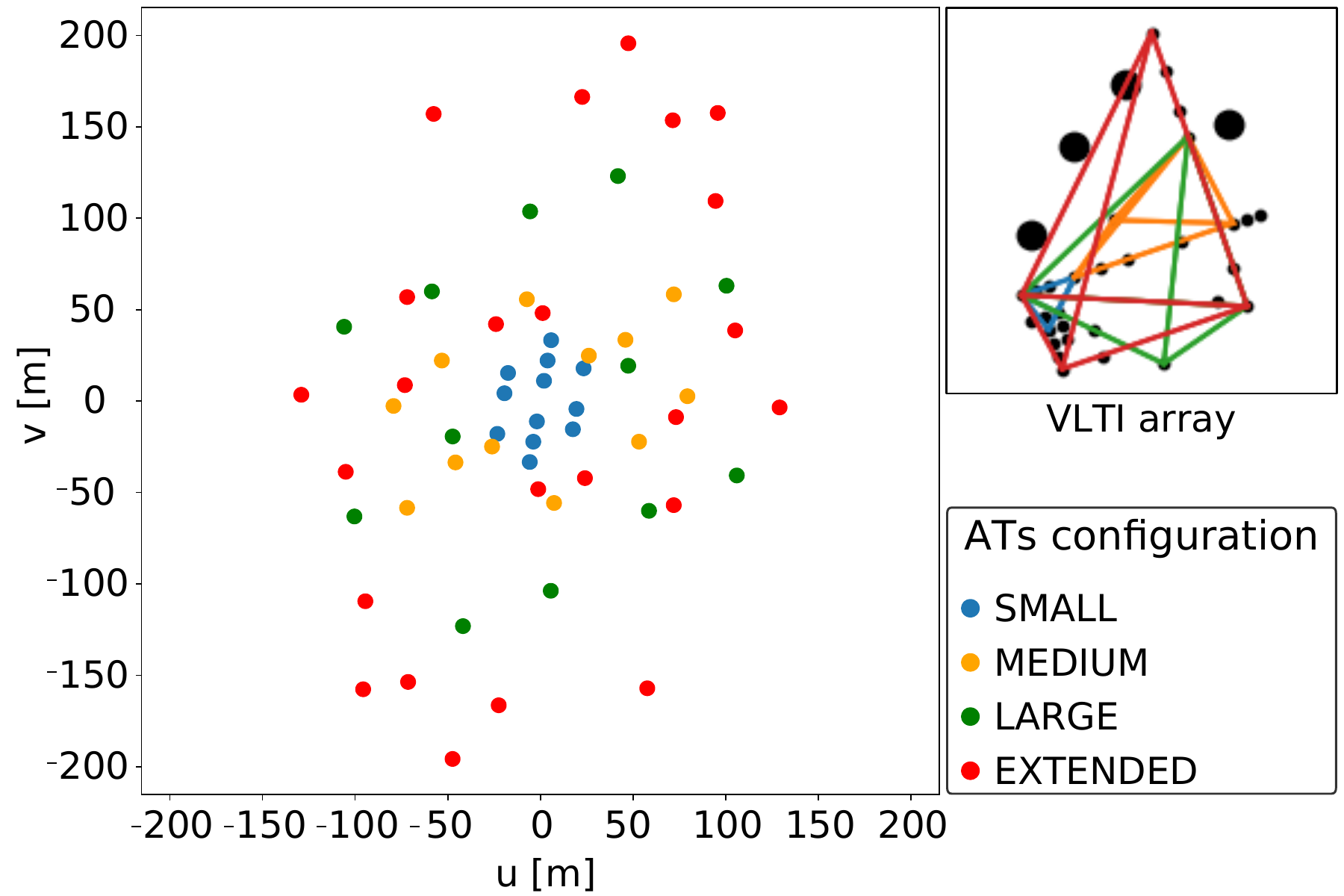}
      \caption{Coverage of the ($u,v$)-plane obtained from the VLTI/MATISSE observations of \object{$3$~Puppis}. The legend in the lower-right corner indicates the colours corresponding to the four AT configurations used: Small (blue), Medium (orange), Large (green), and Extended (red). A schematic map of the VLTI baselines, using the same colour code, is shown in the upper-right corner. East corresponds to increasing $x$-axis values, and North to increasing $y$-axis values.}
         \label{fig:uv} 
  \end{figure}
\section{MATISSE data}\label{sec:obs}

\subsection{VLTI/MATISSE instrument}\label{sec:pres_MATISSE}

The VLTI is the interferometric extension of ESO Very Large Telescope (VLT) located at Paranal Observatory, in Chile. It allows one to coherently combine the light from either the four 8.2\,m Unit Telescopes (UTs) or the four 1.8\,m movable Auxiliary Telescopes (ATs). The Multi AperTure mid-Infrared SpectroScopic Experiment (VLTI/MATISSE, \citealp{MATISSE_Lopez+2022}) is the second-generation mid-IR instrument at the VLTI. It can simultaneously observe in three IR bands --L (3.0–3.9$~\mu$m), M (4.5–5.0$~\mu$m), and N (8.0–13$~\mu$m)-- and provides the scientific community, for the first time, with four-telescope mid-IR imaging capabilities. The two arms of the instrument, dedicated to the LM-bands and the N-band respectively, are equipped with various spectral dispersers offering spectral resolving power ranging from 30 (LOW) to 3300 (HIGH+). MATISSE can operate alone or using the fringe tracker from the VLTI/GRAVITY instrument \citep{Woilez+2024}.

\subsection{$3$~Puppis Observations}\label{sec:obs_MATISSE}
\object{$3$~Pup} was observed three times with MATISSE in February 2020, using baselines ranging from 11 to $130$~m thanks to the Small, Medium and Large configurations available for the movable ATs telescopes. These three observations were then complemented in March 2024 with two additional observations using the newly available Extended ATs configuration giving access to the $200$~m baseline and hence reaching a spatial resolution of $1.5$~mas at $3.0\,\mu$m and $6.2$~mas at $12\,\mu$m (see Table~\ref{table:ATs} for the complete list of baselines). Previous interferometric studies located the inner rim at a radius of $2$~mas in the K-band and $5.7$~mas in the N-band (see Table~\ref{tab:params_prop_lPup_litterature}). These new observations therefore probe the expected sublimation region of the dusty disc and provide improved constraints on its inner rim geometry. Additionally, since no significant temporal variations have been reported in this region, nor detected in interferometric data from the common baseline A0-J2 between 2020 and 2024, we considered that the structural properties have remained stable between the two epochs and justify the merge of datasets for our analysis.\\
As \object{$3$~Pup} is a bright target in each MATISSE spectral band, observations were carried out in stand-alone mode (i.e. without the fringe tracker). As the main purpose of this observation campaign was to focus on continuum emission from 3.0 to 12$~\mu$m, observations were performed using the smallest spectral resolution available (i.e. R$\sim$30) for both LM- and N-bands.\\
The ($u,v$)-plane coverage obtained for \object{$3$~Pup} is plotted in Figure~\ref{fig:uv} and the complete log of the observations can be found in Table~\ref{table:obs_MATISSElog}. Each observing block of the science target \object{$3$~Pup} is preceded and followed by the observation of a calibration star, one dedicated to the LM-bands and the second one for the N-band. A calibration star is an unresolved star or close to unresolved if a precise measurement of its diameter is known. Table~\ref{table:obs_calib} lists the stars used to calibrate our MATISSE observations, as well as their spectral types, their respective angular diameters in L- and N-bands, and their corresponding fluxes.

\subsection{Data Processing}\label{sec:obs_MATISSE_cal}
\begin{figure*}[ht]
   \centering
   \includegraphics[width=0.85\hsize]{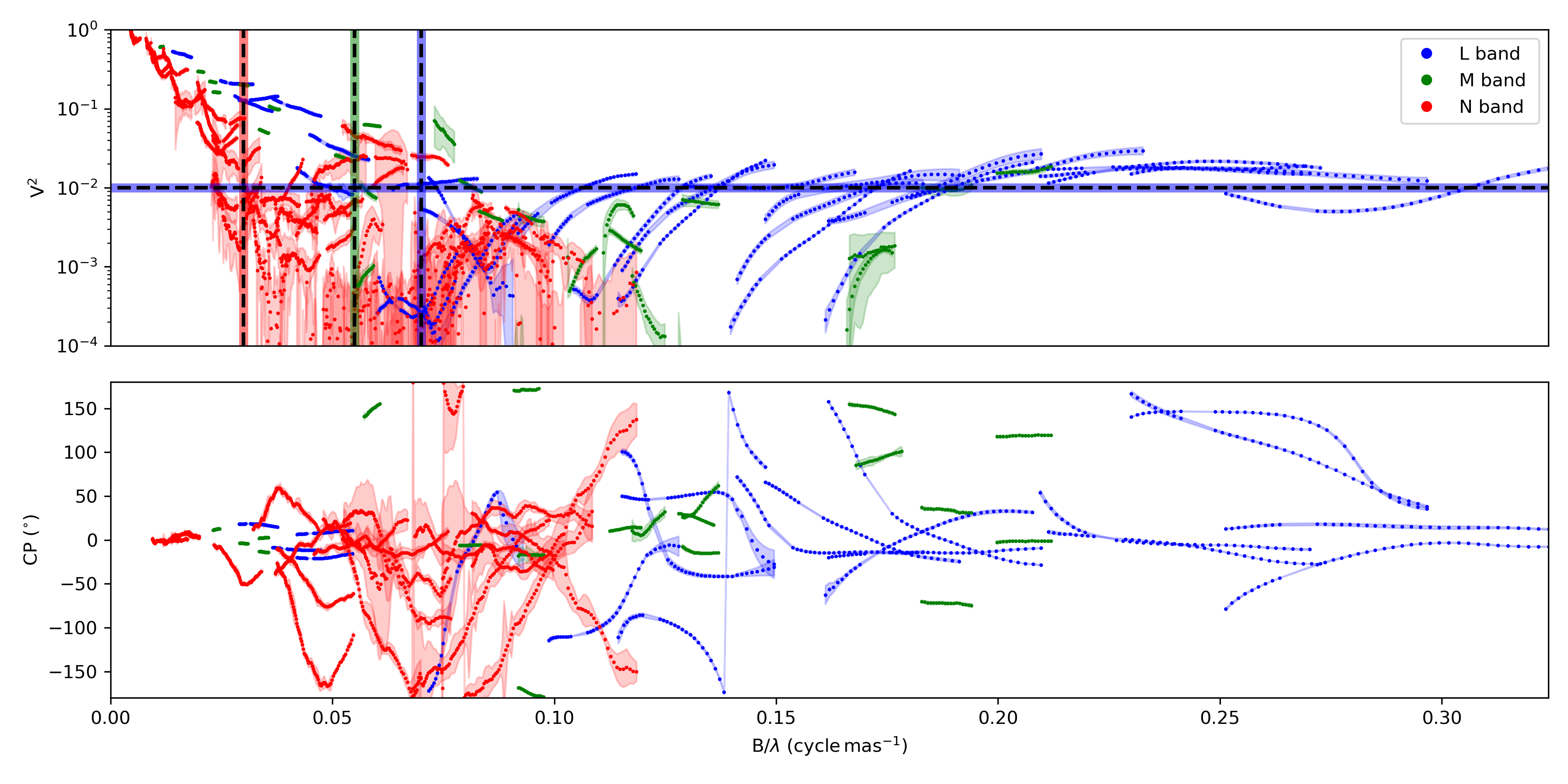}
      \caption{VLTI/MATISSE squared visibility data, denoted V$^2$, in logarithmic scale (\textit{top plot}) and closure phase data, denoted CP, (\textit{bottom plot}) of \object{$3$~Puppis} in the L- (blue), M- (green), and N- (red) bands, plotted as a function of spatial frequency B/$\lambda$. The vertical dashed lines indicate the approximate positions of the first visibility nulls in the three bands with the same colour code. The horizontal dashed line marks the visibility plateau at long baselines in the L-band, which reflects the relative flux contribution from the unresolved central source.}
         \label{fig:dataLMN} 
  \end{figure*}

The MATISSE observations in the L-, M-, and N-bands were reduced using the standard MATISSE Data Reduction Software (DRS) version 2.0.0\footnote{\url{https://www.eso.org/sci/software/pipelines}} \citep{DRS_MATISSE2016}. This pipeline consists of recipes developed by the consortium in the ESO CPL-based reduction framework \texttt{EsoRex} (v3.13.5), plus a set of Python libraries \texttt{mat\_tools}\footnote{\url{https://github.com/Matisse-Consortium/tools}}. The pipelines produced five reduced and calibrated data files in the OIFits format \citep{2017oifits} which contains various interferometric quantities. However, in this paper, we focus on the two quantities needed for the image reconstruction procedure: the squared visibility (V$^2$) and the closure phase (CP) observables. The final reduced and calibrated data for the five observations were then merged into a single file using the \texttt{OiFitsExplorer} tool by the Jean-Marie Mariotti Center (JMMC)\footnote{Available at \url{https://www.jmmc.fr/oifitsexplorer}}. The final L-, M-, and N-band data are plotted as a function of spatial frequency B/$\lambda$ in Figure~\ref{fig:dataLMN}. The spatial frequency quantifies the number of interference fringe cycles that can be resolved by MATISSE within an angular separation of $1$~mas, hence the unit cycles$\cdot$mas$^{-1}$.

\subsection{Qualitative Analysis}\label{sec:obs_qualitative}
From the V$^2$ curves displayed in the top panel of Figure~\ref{fig:dataLMN}, we can estimate the characteristic size of \object{$3$~Pup}'s circumstellar disc by identifying the position of the first visibility null under the uniform disc approximation. In the L-band, the null occurs at approximately $0.07$~cycles$\cdot$mas$^{-1}$, corresponding to a size\footnote{The equivalent spatial extension, or size, can be estimated by taking the inverse of the spatial frequency such as $1/(\mathrm{B}/\lambda) = 1/0.07$~cycles$\cdot\mathrm{mas^{-1}} \sim 14$~mas.} of about $14$~mas. In the N-band, the first null indicates a larger characteristic size of approximately $33$~mas. This clearly demonstrates the increasing extent of the CE from the L-band, which traces warm dust near the sublimation temperature, to the N-band, which is more sensitive to cooler material. The V$^2$ plateau reached at the longest baselines in the L-band provides an estimate of the flux contribution from the compact central source, accounting for approximately $10\,\%$ of the total flux. In contrast, this plateau level cannot be accurately determined from the M- and N-bands data based on the V$^2$ alone.\\
CP curves displayed in the bottom panel exhibit significant variations at spatial frequencies above $0.06$~cycles$\cdot$mas$^{-1}$ in the L- and M-bands, and above $0.04$~cycles$\cdot$mas$^{-1}$ in the N-band. Such variations are commonly observed in circumstellar discs and are typically attributed to the asymmetric (or skewed) aspect of the inner rim of the dusty disc \citep{Monnier2005, Lazareff2017}. However, this effect alone cannot account for the non-zero CPs observed in both L- and M-bands at spatial frequencies lower than those probing the inner rim. Based on estimates from Table~\ref{tab:params_prop_lPup_litterature}, the inner rim lies at a radius of around 5.7\,mas, corresponding to spatial frequencies of approximately 0.08\,cycles$\cdot$mas$^{-1}$. The CP signal of approximately $\pm20\degr$, observed over the spatial frequency range 0.02--$0.05$~cycles$\cdot$mas$^{-1}$ in the L- and M-bands, clearly indicates asymmetries located at higher spatial scales --around 20 to $50$~mas-- well beyond the expected position of the inner rim. Similarly, the CP variations observed in the N-band between 0.025 and $0.05$~cycles$\cdot$mas$^{-1}$ are difficult to reconcile without invoking large-scale asymmetries within the disc structure.\\
Finally, we note that, similarly to the VLTI/MIDI data of \object{$3$~Pup} analysed by \cite{Meilland+2010}, a clear broad silicate emission feature is present in the N-band, leading to a visibility drop around 10~$\mu$m and resulting in a "U"-shaped visibility variation across the band for each partially resolved baseline.

\section{Image reconstructions}\label{sec:method_imaging}
The ($u,v$)-plane coverage obtained from our observations is dense and spatially well distributed (see Figure~\ref{fig:uv}), making these MATISSE interferometric observables suitable for the use of image reconstruction techniques \citep{exhaustiveMiRA_Thiebaut2009}. In this section, we apply two different algorithms to reconstruct images from the MATISSE data. First, we perform a statistical analysis of the images reconstructed using the \texttt{MiRA} software, along with \texttt{PYRA} and \texttt{MYTHRA} which are two new Python-based interfaces for \texttt{MiRA} (\textcolor{blue}{Drevon et al. in prep.}). Then, we use the \texttt{SPARCO} algorithm, specifically developed for reconstructing images of CEs, to compare to the images obtained with \texttt{MiRA}. Both methods offer complementary advantages: \texttt{MYTHRA} enables the production of statistically robust images, while \texttt{SPARCO} enhances image contrast by subtracting the contribution of a bright central source.

\subsection{Statistical image reconstruction procedure}\label{sec:method_statimaging}
This subsection outlines the different steps of the statistical analysis framework performed on the images reconstructed with the \texttt{MiRA} imaging software across all MATISSE spectral channels.

\subsubsection{\texttt{MiRA} imaging tool}\label{sec:method_imaging_MiRA}
The Multi-aperture Image Reconstruction Algorithm (\texttt{MiRA}, \citealp{MiRA_Thiebaut2008SPIE}) is a software based on a Bayesian approach that integrates regularization functions to effectively reconstruct high-resolution images from sparse interferometric data, even under challenging conditions such as limited Fourier phase information. As explained exhaustively by \citet{LeBesnerais+2008} and \citet{MiRA_Thiebaut2008SPIE}, the algorithm follows a gradient-descent method search to minimize the total chi-squared function $\chi^2_\mathrm{tot}$ (a.k.a. the cost or penalty function) as follows:
\begin{equation}
\chi^2_\mathrm{tot} = f_\mathrm{data} + \mu f_\mathrm{prior}\,\, ,
\label{eq:likelihood}
\end{equation}
where $f_\mathrm{data}$ denotes the data fidelity term --measured using the chi-squared statistics\footnote{All $\chi^2$ values reported in this work correspond to reduced chi-squared.}--, $f_\mathrm{prior}$ the regularization term, and $\mu$ a weighting scalar referred to as the hyperparameter that controls the trade-off between them. After testing multiple reconstructions, the hyperbolic regularization, an edge-preserving smoothness prior, was found to be the most suitable regularization function as it outperforms the quadratic criterion, especially for extended objects with sharp features (e.g. disc edge such as an inner rim), while quadratic regularization tends to over-smooth the edges of structures detected within the \object{$3$~Pup} system. \\
Eq.~\ref{eq:hyperbolic_reg} shows how the hyperbolic regularization $f^\mathrm{hyperbolic}_\mathrm{prior}$ is implemented in \texttt{MiRA}. This prior function relies on two tunable parameters: the scale parameter $s$ and the edge-gradient threshold parameter $\tau$, which estimates the typical absolute difference between neighbouring pixels.
\begin{eqnarray}
f^\mathrm{hyperbolic}_\mathrm{prior}(\vec{x})=\tau^2 \sum_{p,q}^{} C\left(\frac{\sqrt{(x_{p+1,q}-x_{p,q})^2+(x_{p,q+1}-x_{p,q})^2}}{\tau s}\right) \label{eq:hyperbolic_reg}\\
C:z \mapsto C(z)=z -\ln(1+z) \quad \mathrm{and} \quad s^2=\frac{1}{\mu}
\end{eqnarray}
with $\vec{x}$ the sought object distribution (i.e. the reconstructed image of the science target), $p$ and $q$ the pixel indexes used to navigate through the dimensions of the sought two-dimension image, and $C$ the cost function. Finally, \texttt{MiRA} requires an initial image as a starting point for the iterative reconstruction process. This initialization can be either a simple parametric model (e.g. uniform disc, limb darkened disc, Gaussian disc) fitted to the data beforehand, or a centred point source (i.e. Dirac delta function). We chose the latter approach as the starting image of each \texttt{MiRA} image to minimize prior assumptions about the object's morphology. 

\subsubsection{Image grid synthesized with \texttt{PYRA}}\label{sec:method_imaging_PYRA}
The IR images are reconstructed using \texttt{PYRA}\footnote{Python for MiRA (\texttt{PYRA}), available at \url{https://github.com/jdrevon/PYRA/tree/main}}, a Python user-friendly wrapper designed to efficiently generate thousands of image reconstructions, enabling extensive exploration of \texttt{MiRA}'s free parameter space. A step-by-step summary of the reconstruction procedure follows.\\
\textbf{Step 1: Image characteristics.} The field of view (FoV) and pixel size of the reconstructed images are determined based on the ($u,v$)-plane coverage (see Figure~\ref{fig:uv}). From this coverage, two key quantities are extracted: $B_\mathrm{max}$ and $B_\mathrm{min}$, corresponding respectively to the longest and the shortest projected baselines used in the observation of \object{$3$~Pup}; see Table~\ref{table:ATs} for the values. The pixel size is estimated using the lowest angular size resolved by MATISSE during \object{$3$~Pup} observations, noted $\theta_\mathrm{r}$, at a given wavelength\footnote{Here we chose the centred wavelength for a given spectral range.} $\lambda_0$, and is defined as follows:
\begin{eqnarray}\label{eq:angular_resolution_MATISSE}
\theta_\mathrm{r} = \frac{\lambda_{0}}{2 B_\mathrm{max}} \,\, .
\end{eqnarray}
However, due to the sampling criteria, the pixel size of the resulting synthetic image $\theta^+$, should be equal to half the smallest angular resolution of the instrument \citep{exhaustiveMiRA_Thiebaut2009}. The physical FoV of the reconstructed image is eventually determined by the interferometric FoV, instead of the photometric FoV, which following \cite{GALARIO_Tazzari+2018} is given by:
\begin{eqnarray}\label{eq:FoV_MATISSE}
\mathrm{FoV}_\mathrm{interf} =\frac{\lambda_{0}}{B_\mathrm{min}}\,\, . 
\end{eqnarray}
\noindent \textbf{Step 2: Parameters sampling.} To ensure that \texttt{PYRA} probes efficiently both the FoV and the pixel size parameter spaces, seven different FoV values (= $N_\mathrm{FoV grid}$) and four different pixel size (= $N_\mathrm{scale grid}$) values are randomly taken from the following intervals:
\begin{eqnarray}\label{eq:grid_FoV}
\mathrm{FoV}_\mathrm{grid} &=& \overbrace{[0.5 \times\,\mathrm{FoV}_\mathrm{interf}\,,\ldots\,, 1.5\times\,\mathrm{FoV}_\mathrm{interf}]}^\text{7 values}\, ,\\
\theta_\mathrm{grid} &=& \underbrace{[\theta^+\,,\ldots\,, \theta_\mathrm{r}]}_\text{4 values}\,\, \label{eq:grid_pixelscale}.
\end{eqnarray}
Table~\ref{tab:resolution_band} informs of the values obtained for the angular resolution of MATISSE and the physical FoV for each spectral band covered.
Then, for each combination of FoV and pixel size values, we establish a grid of values for the regularization parameter of the hyperbolic prior function ($\tau_\mathrm{grid}$). Since $\tau$ depends on both pixel size and FoV quantities (see Eq.~\ref{eq:tau}), the grid is reinitialized whenever either variable changes within a given spectral band. As a result, the threshold parameter is systematically adapted to the spatial characteristics of the reconstructed image. The six values of $\tau$ (= $N_{\tau\mathrm{ grid}}$) are chosen logarithmically spaced and computed as:
\begin{eqnarray}
\forall (k,l) \in [\![1,4]\!] \times [\![1,7]\!],\, \tau^{k,l} = \left(\frac{\theta_\mathrm{grid}^k}{\mathrm{FoV}_\mathrm{grid}^l}\right)^2\label{eq:tau} \\
\mathrm{hence} \quad \tau_\mathrm{grid}^{k,l} = \overbrace{\left[\frac{\tau^{k,l}}{10^4}\,,\frac{\tau^{k,l}}{10^3}\,, \frac{\tau^{k,l}}{10^2}\,, \frac{\tau^{k,l}}{10}\,, \tau^{k,l}\,, 10\times\tau^{k,l} \right]}^\text{6 values}\label{eq:grid_tau_MiRA}
\end{eqnarray}
The index $k$ refers to the $k^\mathrm{th}$ component of the pixel scale grid ($\theta_\mathrm{grid}$, see Eq.~\ref{eq:grid_pixelscale}), while $l$ denotes the $l^\mathrm{th}$ component of the FoV grid (FoV$_\mathrm{grid}$, see Eq.~\ref{eq:grid_FoV}), both defined for each spectral band.\\
\textbf{Step 3: Hyperparameter.} To identify the optimal hyperparameter value and produce the best possible reconstructed image, we perform a broad logarithmic search with \texttt{PYRA} across the interval $[10^0\,,10^8]$. The sampling strategy of the hyperparameter grid consists in generating randomly twenty values of $\mu$ (= $N_{\mu\mathrm{ grid}}$), gathered under the denoted $\mu_\mathrm{grid}$, within the mentioned interval. However, these $\mu$ values are reinitialized for every $\tau$ value explored in $\tau_\mathrm{grid}^{k,l}$ (see Eq.~\ref{eq:grid_tau_MiRA}), ensuring that each $\tau$ uses a new sampling of $\mu_\mathrm{grid}$, leading to a comprehensive exploration of the hyperparameter space rather than a fixed linear grid and preventing also redundant sampling. In other words, for every $\tau$ value, we explore a full different set of N$_\mathrm{\mu grid}$ values of $\mu$.\\
\textbf{Step 4: Iteration threshold.} The remaining \texttt{MiRA} variables are kept at their default values in \texttt{PYRA}: scale of finite differences $\eta = 1$, tolerance factors $f_\mathrm{tol} = g_\mathrm{tol} = x_\mathrm{tol} = 0$, and total flux normalisation $f_\mathrm{tot} = 1$. The maximum number of iterations (or evaluations) is set to 1000, following the findings of \textcolor{blue}{Drevon et al.~(2025, in prep.)}, which suggests this number as a reasonable limit to allow \texttt{MiRA} to converge on an image with a low $\chi^2$ value while minimizing the impact of artefacts. \\
\textbf{Step 5: Grid synthesis.} With all the aforementioned variables set, \texttt{PYRA} is run by systematically exploring a grid of each combination of $\mu$, $\tau$, FoV and pixel size values. For each spectral band sought, it results in a grid of images with a total of $N_\mathrm{MiRAgrid} = 3\,360$ different reconstructed images\footnote{$N_\mathrm{MiRAgrid} = N_{\mu\mathrm{grid}} \times N_\mathrm{FoVgrid} \times N_\mathrm{scalegrid} \times N_{\tau\mathrm{grid}} = 20 \times 7 \times 4 \times 6$.}$^{,}$\footnote{The 3360-image grid was computed using a 12th Gen Intel(R) Core(TM) i7-12800H processor with a 15 GB of RAM. Under these computational conditions, the grid required different processing times depending on the spectral channel: 8h37min for the L-band, 3h35min for the M-band, and 21h35min for the N-band.}.

\subsubsection{Averaged images computed with \texttt{MYTHRA}}\label{sec:method_imaging_statanalysis}
Once \texttt{PYRA} has generated the grid of synthetic images, \texttt{MYTHRA}\footnote{Mean Astrophysical Images with PYRA (\texttt{MYTHRA}), available at \url{https://github.com/jdrevon/MYTHRA/tree/main}} is used to derive a final image for each spectral band. To do so, \texttt{MYTHRA} analyses the \texttt{PYRA} grid by identifying the optimal hyperparameter value, denoted $\mu^+$, based on an L-curve diagram. By definition an L-curve plots the resulting chi-squared ($\chi^2$) value of each synthetic image in the \texttt{PYRA} grid as a function of its associated hyperparameter value. This graphical representation helps determine the $\mu^+$ value because it corresponds to the inflection point of this curve. This point is located just before $\chi^2$ begins to diverge to a plateau \citep{Dalla_Vedova+2017} and \texttt{MYTHRA} estimates it using a gradient-based search. As $\mu$ increases, the influence of the data in the minimization process decreases, leading to an increase in $\chi^2$. The set $\mu^+$ value therefore marks the transition at which the regularization term begins to dominate and degrade the model's fit to the data. The L-curves obtained in the L-, M-, and N-bands for \object{$3$~Pup} MATISSE data, along with their corresponding $\mu^+$ values, are shown in Figure~\ref{fig:Lcurve}. Then \texttt{MYTHRA} is able to establish a robust subset of the \texttt{PYRA} grid by keeping only the reconstructed images in the vicinity of the inflection point to exclude outliers. The $\chi^2$ criterion is also employed to refine the selection of the final subset by constraining the vertical axis of the L-curve plot. After the identification of a statistically reliable subset of reconstructed images (i.e. combining the $\mu^+$ and $\chi^2$ criteria), \texttt{MYTHRA} resamples all the selected images to a common pixel size and spatial resolution --the rest of \texttt{MiRA} variables are kept identical otherwise.\\
Before producing the final averaged image, \texttt{MYTHRA} applies an outlier rejection step to refine one last time the subset. This is done through an iterative process in which images are added one by one to a cumulative mean. Therefore, at each iteration, if the inclusion of a new image causes the global $\chi^2$ to exceed a predefined threshold ($\chi_\mathrm{global}^2 = 10$ in this work), the current image is then excluded from the final subset. The resulting `cleaned' subset is used to eventually compute the final averaged image. The final averaged image for each spectral band studied are presented in the first row of Figure~\ref{fig:images}. The corresponding (simulated) V$^2$ and CPs extracted from the averaged image, shown in Figure~\ref{fig:fit_LMNband}, indicate a satisfactory fit to the MATISSE data given the complexity of the datasets, with global $\chi^2$ values equal to 7.6, 5.6, and 2.2 for the L-, M-, and N-bands respectively. Table~\ref{tab:outputs_MYTHRA} summarizes the \texttt{MYTHRA} statistics obtained at each step to produce the averaged images. 

\subsection{\texttt{SPARCO} imaging tool}
The combination of \texttt{MiRA} with the \texttt{PYRA} and \texttt{MYTHRA} interfaces enables the reconstruction of statistically relevant images. However, in the context of CEs, the dynamic range of a reconstructed image can be significantly limited by the presence of a bright, central point-like source (i.e. the unresolved component) --in our work, it refers to the combined emission from the binary system and the gaseous disc. To address this limitation, we complemented our imaging analysis with the Semi-Parametric Approach for Reconstruction of Chromatic Objects (\texttt{SPARCO}, \citealp{SPARCO_Kluska+2014}), an algorithm specifically designed to overcome such challenges.\\
In \texttt{SPARCO}, the optical interferometric data are modelled as the sum of a geometrical model (generally a chromatic unresolved object) plus a reconstructed image, each with their own spectrum. Within this framework, the image is reconstructed using the \texttt{MiRA} algorithm (along with its specific parameters), while the model component is described using additional parameters. We used the implementation of the \texttt{SPARCO} algorithm in the \texttt{OImaging} service\footnote{Available at \url{https://www.jmmc.fr/oimaging}} by the JMMC which provides user-friendly access to several interferometric image reconstruction algorithms. The \texttt{SPARCO} images presented in the bottom row of Figure~\ref{fig:images} have been reconstructed separately on the L-, M-, and N-bands, and with the following set parameters: a reference wavelength set at 3.0$~\mu$m; a black-body profile with a temperature of 1\,500\,K for the \texttt{MIRA} image initialised with a centred Gaussian disc; and a point-source object with a flux ratio 0.12 (except for the M-band which is equal to 0.15) and a temperature of $10\,000$~K for the geometric model.

\subsection{Comparison between \texttt{MYTHRA} and \texttt{SPARCO} images}\label{subsec:MYTHRA_SPARCO_comp}
\begin{figure*}[h]
\centering
\includegraphics[width=0.8\hsize]{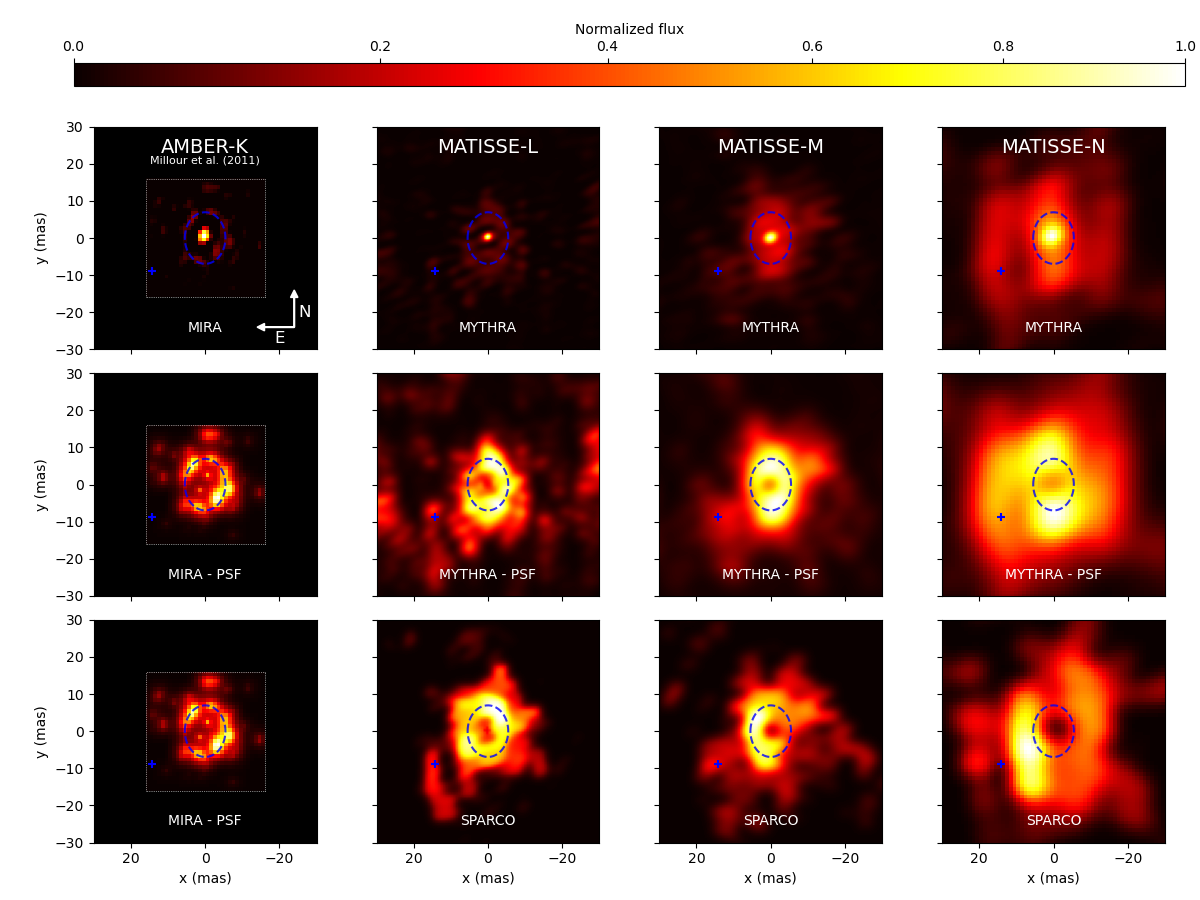}
  \caption{Final image reconstructions of $3$~Puppis in the mid-infrared using \texttt{MiRA}, \texttt{MYTHRA}, and \texttt{SPARCO} imaging tools. The field of view for each image is (60 mas $\times$ 60 mas) with North up and East left. The blue dashed ellipse indicates the best-fit inner rim of the dusty disc derived from VLTI/AMBER data, while the blue cross marks the southeastern elongated clump best-fit position, denoted $R_\mathrm{ref}$, from MATISSE L-band geometric modelling (at a distance of $16.71 \pm 0.03$~mas from the image centre). Brightness colour scale is normalised to peak intensity (i.e. maximum pixel value). \textit{K-band image:} The left-most column shows the images obtained with VLTI/AMBER data by \cite{Millour+2011}. In the first row the median \texttt{MiRA} image is displayed, and the last two rows present the identical convolved median \texttt{MiRA} image with the $\lambda/2B_\mathrm{max}$ PSF subtraction applied. \textit{L-M-N-bands images:} The remaining three columns display the images obtained with VLTI/MATISSE. On the first row the resulting \texttt{MYTHRA} averaged image is shown. The second row shows the convolved image of \texttt{MYTHRA} with the $\lambda/2B_\mathrm{max}$ PSF subtraction applied. The last row gives the resulting \texttt{SPARCO} image, convolved with the interferometric beam.}
  \label{fig:images}
\end{figure*}

To enable a direct comparison between \texttt{SPARCO} and \texttt{MYTHRA} images in each spectral band of MATISSE, we applied the Point Spread Function (PSF) subtraction technique on the \texttt{MYTHRA} reconstructions (\texttt{MYTHRA} $-$ PSF). The technique begins with the averaged image produced by \texttt{MYTHRA} (see Sect.~\ref{sec:method_imaging_statanalysis}). From this unconvolved image we subtract a Gaussian PSF centred on the brightest pixel and scaled to match its maximum intensity. The full width at half maximum (FWHM) of the PSF is equal to $\theta_\mathrm{r}$ as defined in Eq.~\ref{eq:angular_resolution_MATISSE} (see the values for each band in Table~\ref{tab:resolution_band}). The residual image is then convolved with the interferometric beam characterized in Table~\ref{tab:interferometric_beam}). The resulting image from this operation is denoted as the \texttt{MYTHRA} - PSF image and is shown in the second row of Figure~\ref{fig:images}. For consistency, the same procedure was also applied to the VLTI/AMBER K-band image from \citet{Millour+2011}, which is also displayed in the same figure.\\
The L-band images reconstructed with the imaging tools \texttt{MYTHRA} and \texttt{SPARCO} are in excellent agreement, both revealing a skewed inner rim with the same orientation of asymmetry and similar contrast. In addition, an elongated feature located approximately 20\,mas southeast of the central object is clearly visible in both reconstructions --hereafter referred to as the southeastern (SE) elongated clump. This structure is also present in the M-band images, although with lower contrast.\\
Additionally, a second structure seems to appear in the northwest (NW) region of M-band images, especially for \texttt{SPARCO} and \texttt{MYTHRA} $-$ PSF. Both exhibit a skewed inner rim; however, in the case of the \texttt{SPARCO} image the orientation of the skewness is reversed, which is unlikely to reflect a physical property of the system. The N-band images likewise suggest the presence of the eastern structure and a fainter one to the west. While the overall sizes are comparable, \texttt{SPARCO} images appear smoother, with fewer artefacts, than those produced by \texttt{MYTHRA}.
\begin{figure*}[h]
   \centering
   \includegraphics[width=0.7\hsize]{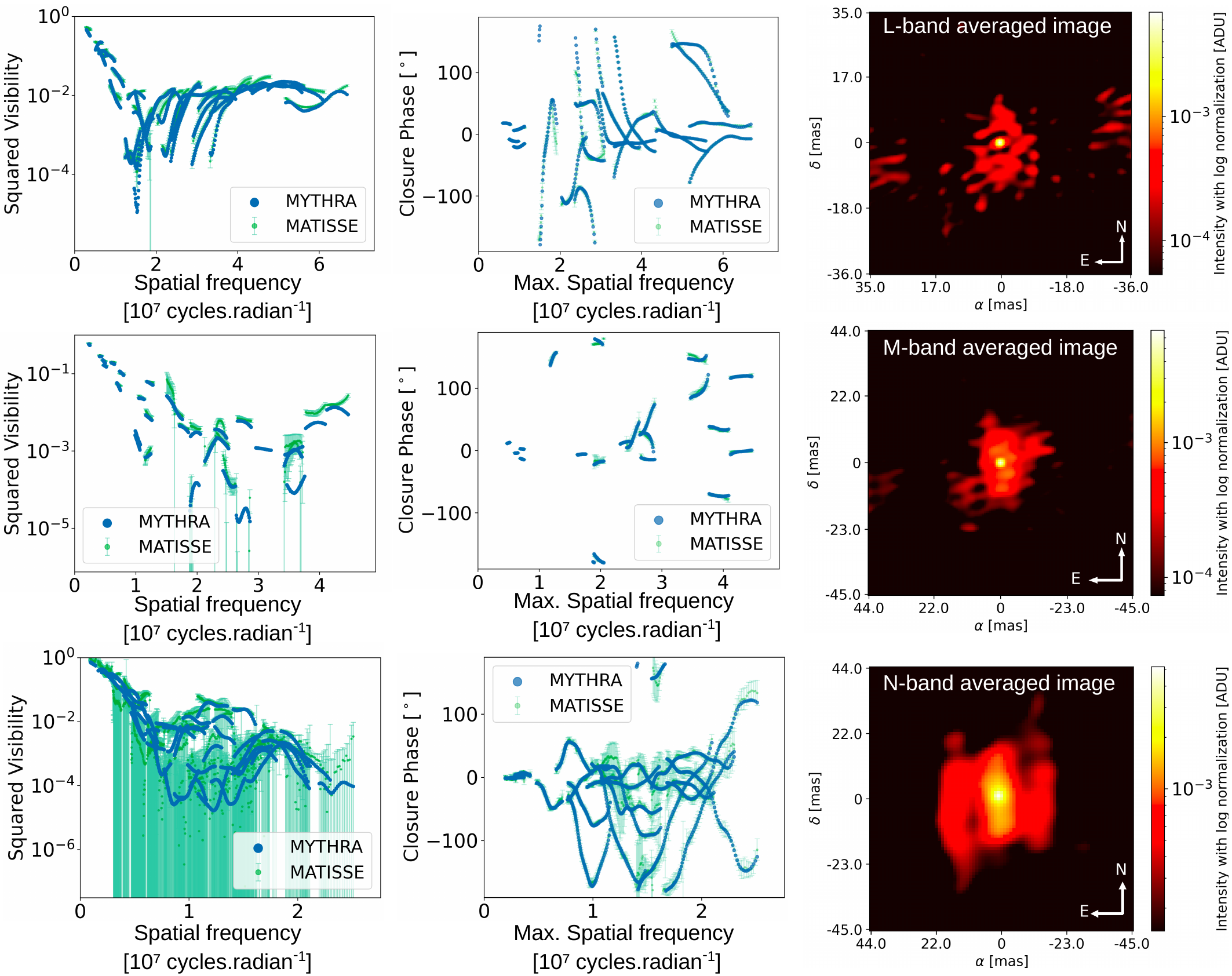}
  \caption{Comparison between the interferometric observables extracted from the \texttt{MYTHRA} averaged images (in blue) and the VLTI/MATISSE data of \object{$3$~Puppis} (in green), shown respectively in the L-band (\textit{first row}), in the M-band (\textit{second row}), and in the N-band (\textit{third row}). \textit{Left column:} Squared visibility amplitudes as a function of spatial frequency in logarithmic scale. \textit{Central column:} Closure phases as a function of spatial frequency, computed using the longest baseline. \textit{Right column:} Final averaged images resulting from \texttt{MYTHRA}, normalized to unity (i.e. the sum over all pixels equals one), with a logarithmic colour scaling, and not convolved with the interferometric beam. The minimum value for each brightness scaling is set to half the standard deviation of the associated image to filter out the artefact produced by the \texttt{MiRA} algorithm.} \label{fig:fit_LMNband}
\end{figure*}

\section{Discussion}\label{sec:discuss}

\subsection{Large-scale asymmetries detected in the disc} \label{subsec:discussion_asymmetry}
The reconstructed images allow us to identify the main spatial components of \object{$3$~Pup}'s CE and thereby characterize the general morphology of its dusty disc across different mid-IR spectral bands. Thanks to the statistical image reconstruction procedure presented in Sect.~\ref{sec:method_statimaging}, the final images plotted in Figure~\ref{fig:images} reveal a large-scale brightness distribution asymmetry in the SE region of the CE in the L-band (see Sect.~\ref{subsec:MYTHRA_SPARCO_comp}). This asymmetry, which we refer to as an elongated clump, seems to be responsible for the non-zero CP signal at small spatial frequencies in the MATISSE data (i.e. around 0.02--0.05\,cycles$\cdot$mas$^{-1}$), as displayed in the bottom panel of Figure~\ref{fig:dataLMN}. \\
Therefore, our findings demonstrate that the CE of \object{$3$~Pup} cannot be described by a simple disc with a skewed inner rim alone since at least one bright asymmetric structure is consistently detected in all IR images, at approximately 20\,mas SE of the image centre --position marked by the blue cross in Figure~\ref{fig:images} and as determined in Sect.~\ref{subsec:geomodel_asymmetry}. The presence of this elongated clump is also detected in the M- and N-band images as it appears at a similar SE location from the image centre. Moreover, another (fainter) asymmetry in the NW region seems to be visible in the M- and N-bands images but absent from the L-band image.
\subsection{Geometric modelling of the L-band asymmetry} \label{subsec:geomodel_asymmetry}
To constrain quantitatively the position of the asymmetric feature identified in the L-band images as the SE elongated clump, we used \texttt{oimodeler}\footnote{Available at \url{https://oimodeler.readthedocs.io}} \citep{2024MeillandOimodeler}, a modular modelling tool for optical interferometry which includes a Markov Chain Monte Carlo sampling algorithm (\texttt{emcee}, \citealp{emcee_2013}). The fit was performed in the L-band only and limited to low spatial frequencies (i.e. baselines shorter than 40\,m) to minimize contamination from smaller-scale structures. The central object was modelled as a centred Gaussian, while the asymmetry was modelled as a second Gaussian component. The fit yields a position of the asymmetry at $x=14.21 \pm 0.03$~mas and $y=-8.79 \pm 0.04$~mas, or a separation with the central source of $R_\mathrm{ref} = 16.71 \pm 0.03$~mas with a $121.7\degr$ orientation from the north to the east (a.k.a. the position angle, denoted $\mathrm{PA}$). \\
Such values are consistent with the early qualitative analysis made in Sect~\ref{sec:obs_qualitative} and the observations made in  Sect~\ref{subsec:MYTHRA_SPARCO_comp} based on the reconstructed images shown in Figure~\ref{fig:images} where this fitted SE asymmetry position is indicated by a blue cross. Converting this angular separation $R_\mathrm{ref}$ to physical units, we obtain a projected physical distance of approximately $R_\mathrm{ref} = 16.71\cdot 10^{-3} \times 631 \sim 10$~au from the \object{$3$~Pup} system centre at the adopted distance $d_\mathrm{GeDR2}$ ($R_\mathrm{ref}\sim 17$~au at $d_\mathrm{PGeDR3}$ respectively). 

\subsection{Comparison to previously published VLTI data}\label{subsec:discuss_compAMBERMIDI}
\subsubsection{VLTI/AMBER K-band data}\label{subsec:discuss_compAMBER}
\object{$3$~Pup} was previously observed in the near-IR during the first semester of 2010 using the VLTI/AMBER instrument. The observations were carried out at high spectral resolution (R$\sim$12\,000), centred on the prominent Br$\gamma$ emission line. The data, published in \citet{Millour+2011}, allowed the authors to constrain the geometry of the object in the near-IR continuum --revealing a compact source surrounded by a skewed inner rim of the dusty disc-- but also the geometry and dynamics of the inner gaseous disc, which was found to be dominated by Keplerian rotation. The image reconstruction of the K-band data was performed using \texttt{MiRA} and a self-calibration algorithm, producing narrow-band images across the Br$\gamma$ line and the adjacent continuum.\\
The data were modelled using a skewed ring to represent the inner rim of the dusty disc, along with a kinematic model for the line-emitting gas. They found a diameter of 13.9\,mas for the dusty inner rim and an elongation ratio (a.k.a. flattening) of 1.27, which corresponds to an inclination angle of $38\degr$ under the assumption of a geometrically thin disc. These estimates are consistent with the first null at 0.07 cycles$\cdot$mas$^{-1}$ qualitatively estimated in Sect.~\ref{sec:obs_qualitative}. The K-band median AMBER image, along with its PSF-subtracted version, are displayed in the left column of Figure~\ref{fig:images}. The best-fit inner rim model from \citet{Millour+2011} is overplotted over the reconstructed K-, L-, M-, and N-band images. The size, flattening, PA, and skewness of the dusty inner rim derived from the AMBER data (see Table~\ref{tab:params_prop_lPup_litterature}) are fully consistent with our MATISSE reconstructions in the L- and M-bands, shown in the same figure. \\
However, we note that the bright SE elongated clump observed in the MATISSE images as detailed in the previous Sects.~\ref{subsec:discussion_asymmetry} and \ref{subsec:geomodel_asymmetry} is not present in the AMBER data. This discrepancy may be due to the limited FoV used for the AMBER image reconstruction, which was restricted to 32\,mas, while in the L-band MATISSE image instead was reconstructed with a FoV equal to 63\,mas.

\subsubsection{VLTI/MIDI N-band data}\label{subsec:discuss_compMIDI}
\object{$3$~Pup} was also observed in the N-band with the VLTI/MIDI instrument by \cite{Meilland+2010} between October 2006 and January 2008. The authors acquired data on nine baselines ranging from 13 to 71\,m, and determined the wavelength-dependent extension of the CE by fitting a two-component model consisting of an elliptical Gaussian distribution and an extended background.
\begin{figure}[ht]
\centering
\includegraphics[width=0.9\linewidth]{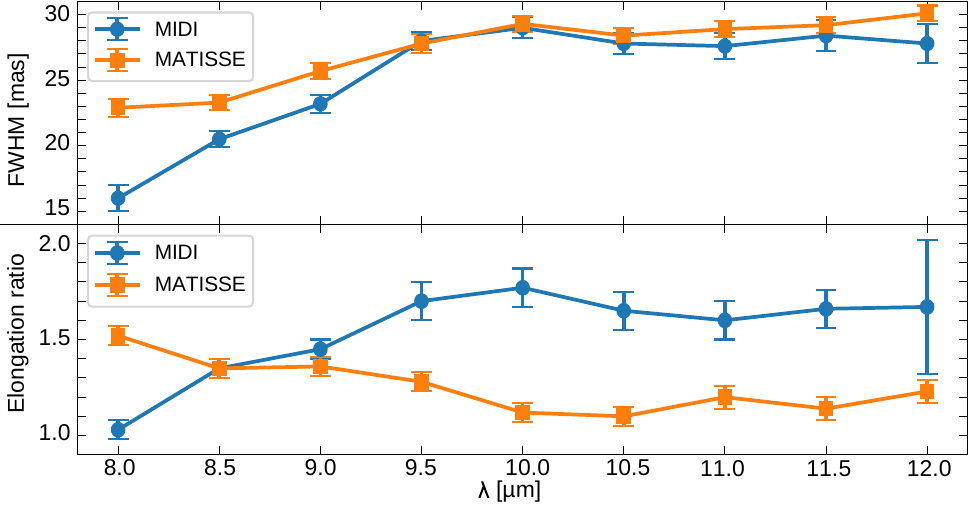}
\caption{Spectral evolution of the fitted disc parameters from a chromatic two-component geometric model applied to \object{$3$~Puppis} N-band observations obtained with the VLTI/MIDI (in blue) and VLTI/MATISSE (in orange) instruments. \textit{Top plot:} The full width at half maximum (FWHM) of the major-axis as a function of wavelength. \textit{Bottom plot:} Elongation (or flattening) ratio as a function of wavelength.}
\label{fig:FWHM_skw_MIDIandMATISSEmodel}
\end{figure}
To enable a direct comparison between the MIDI and MATISSE datasets, since they both observed in the N-band, we restricted the analysis to V$^2$ data only, as CPs were excluded because MIDI was a two-telescope beam-combiner instrument at the VLTI \citep{MIDI_Leinert+2003}. By comparing the V$^2$ data from MATISSE to MIDI, we find that they are in qualitative agreement as shown in Figure~\ref{fig:MIDIandMATISSEdata}. \\
For a quantitative comparison between the MIDI and MATISSE datasets, we kept from the MATISSE V$^2$ data only measurements with baselines shorter than 70\,m to be consistent. Then, we fitted the same chromatic two-component geometric model, consisting of an elongated Gaussian disc distribution and a fully unresolved component (i.e. point source function), to each dataset independently using \texttt{oimodeler}. Two physical parameters have been set as chromatic quantities: the major-axis FWHM of the Gaussian distribution's major axis and the elongation ratio of the disc. The chromatic plot of these derived quantities is shown in Figure~\ref{fig:FWHM_skw_MIDIandMATISSEmodel}. \\
We note that while the FWHM values are consistent in the 9.2–12$~\mu$m wavelength range (ranging from 28 to 30\,mas) for both instruments, a significant discrepancy is observed at shorter wavelengths. Specifically, the MIDI data show a drop in the FWHM to 16\,mas at 8$~\mu$m, whereas the MATISSE data yield a larger value of 23\,mas. Another key difference lies in the elongation ratio, which is systematically higher for MIDI (average of 1.54) than for MATISSE (average of 1.23), and follows a different trend as a function of wavelength. A possible explanation for the reduced flattening is the effect of the asymmetries, which tend to artificially enhance the eastwest expansion of the disc in the N-band, as seen in the reconstructed images in Figure~\ref{fig:images}.\\
Concerning the remaining quantities of the Gaussian component of the geometric model fitting that have been assumed achromatic, we obtain for the MIDI data a PA of $+22\pm59\degr$ and a relative flux of $77\pm3\,\%$, while for the MATISSE data we find a PA of $-14\pm1 \degr$ and a relative flux of $83\pm1\,\%$. Table~\ref{table:Nband_geometricmodel} summarizes the values derived for fitted parameters of the chromatic geometric model. 

\subsubsection{Radiative transfer model with \texttt{MC3D}}\label{subsec:discuss_mc3d}
\begin{figure}[ht]
\centering
\includegraphics[width=0.78\linewidth]{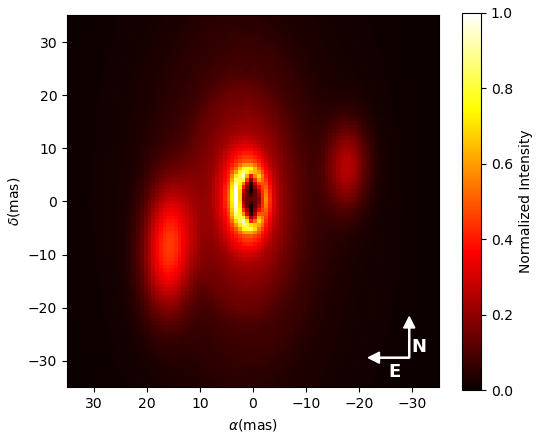}
\caption{Best-fit image at 10$~\mu$m using the N-band MATISSE data of \object{$3$~Puppis}. The fit is performed combining \texttt{MC3D} code with two elliptical Gaussian disc models (\{MC3D + 2EG\} model) to describe the asymmetries revealed by \texttt{MYTHRA} and \texttt{SPARCO} image reconstructions. The model fitting results in a reduced chi-squared value of 24. East corresponds to increasing $x$-axis values, and North to increasing $y$-axis values.} 
\label{fig:MC3D_Egauss}
\end{figure}

Moreover, the authors of \citet{Meilland+2010} also compared the MIDI data to radiative transfer models computed for instance with the \texttt{MC3D} code \citep{MC3D_Wolf+1999,MC3Dv2_Wolf2003}. Although their model was fitted using sparse data with significantly shorter baselines and no CP information, we decided to compare the MATISSE N-band data to synthetic $V^2$ and CPs generated from their best-fit Keplerian-disc model (referred to as the \texttt{MC3D} model). As shown in Figure~\ref{fig:MATISSE_M3CDmodel} with the synthetic interferometric observables displayed in red, this model poorly reproduces the MATISSE data, yielding a reduced $\chi^2$ of 117. While the Keplerian-disc model qualitatively fits the overall size and flattening of \object{$3$~Pup} MATISSE data in the N-band, it fails to reproduce the asymmetries associated to the non-zero CP signal. However, the inability to fit the CPs at frequencies below those of the inner rim was expected as previously mentioned in Sect.~\ref{sec:obs_qualitative}. \\
We tried to minimize the reduced $\chi^2$ value by adding to the original \texttt{MC3D} model an elongated Gaussian component (i.e. \{MC3D + EG\} model) to account for the SE asymmetry revealed in MATISSE data (see Sect.~\ref{subsec:discussion_asymmetry}). This second model fitting leaded to an improved $\chi^2$ value of 47. \\
Thanks to this successful attempt, we tried also to add a second elongated Gaussian component, defined as the \{MC3D + 2EG\} composite model, to account for the fainter NW asymmetry visible in the M- and N-band images. Finally we obtained an even better agreement than the two previous models, with a reduced $\chi^2$ value equal to 24, and achieved a qualitative fit of the CP signature at short spatial frequencies as shown overplotted on the same Figure~\ref{fig:MATISSE_M3CDmodel}, in blue. The best-fit parameters values for the two elliptical Gaussians of the \{MC3D + 2EG\} model are given in Table~\ref{table:MC3D+2EG} and the corresponding image of the composite model is shown in Figure~\ref{fig:MC3D_Egauss}. From the fitted Cartesian coordinates, we compute for each elliptical Gaussian component the radius $R$ from the image centre. We obtain $R_\mathrm{SE}=18.1 \pm 1.8$~mas and $R_\mathrm{NW}=19.0 \pm 2.3$~mas for the SE and NW components respectively. The best-fit position of the brighter asymmetry (i.e. SE elongated clump) is consistent with the values derived in Sect.~\ref{subsec:geomodel_asymmetry}, and the latter contributes to about $13\%$ of the total flux at $10\,\mu$m. 
\begin{table}[ht]
\caption{\label{table:MC3D+2EG} Best-fit parameters for the two elliptical Gaussian disc components from the \{MC3D + 2EG\} model.} 
\centering
\begin{tabular}{c c c c c c c}
\hline \hline
\noalign{\smallskip}
EG & $x$ & $y$ &$\mathrm{PA}$& $e$& $f$& {\small FWHM}\\
comp.& (mas) & (mas)& ($\degr$) & (...) & ($\%$) &(mas)\\
\hline
\noalign{\smallskip}
SE& $+15.6$& $-9.1$& $+1.5$ & $2.6$ & $13$ & $19.1$\\
\noalign{\smallskip}
NW& $-18.0$& $-6.1$& $-1.1$ & $2.1$ & $4.0$ & $13.2$\\
\noalign{\smallskip}
\hline \hline
\noalign{\smallskip}
EG & $\Delta x$ & $\Delta y$& $\Delta \mathrm{PA}$& $\Delta e$& $\Delta f$& $\Delta$~{\small FWHM}\\
comp.& (mas) & (mas)&  ($\degr$) & (...) & ($\%$) &(mas)\\
\hline
\noalign{\smallskip}
SE& 1.2&2.9 &12.8 &2.0  &6.0 &3.6 \\
\noalign{\smallskip}
NW&2.2 &2.7 & 24.7 & 4.0 &4.0  &5.8 \\
\noalign{\smallskip}
\hline
\end{tabular}
\tablefoot{The best fit model to the VLTI/MATISSE data yielded to a reduced chi-squared of 24. For each elliptical Gaussian disc component structure (EG comp.), the free-parameters are ($x$, $y$), PA, $e$, $f$ and FWHM. They stand for the disc's Cartesian coordinates on sky, its major-axis' position angle, its elongation ratio, its flux contribution to the total flux in the image, and its major-axis full width at half maximum size respectively. The upper table gives the averaged fitted value and the lower table gives the corresponding uncertainties for the fitted parameters (denoted with the prefix `$\Delta$' applied to the parameter symbol). `SE' and `NW' acronyms correspond to the southeastern and northwestern asymmetries respectively detected in \object{$3$~Puppis}' dusty disc.}
\end{table}

\subsection{Possible origins for the asymmetric structures} \label{subsec:discussion_origins}
The origin of the asymmetries observed in the mid-IR images of \object{$3$~Pup} remains uncertain. However, \object{$3$~Pup} is a confirmed binary system composed of two evolved stars: a SG primary and a dwarf companion, separated by $1.11$~au, with estimated masses of $8.8 \pm 0.5\,M_\sun$ and $0.75 \pm 0.25\,M_\sun$ respectively, as recalled in Table~\ref{tab:params_prop_lPup_litterature}. \citet{Miroshnichenko+2020} proposed that the circumbinary disc likely formed as a result of common binary evolution involving mass transfer between the components, with part of the transferred material being ejected into the circumstellar environment. We consider four main hypotheses (Hyps.) to explain the observed elongated structures SE and NW as defined in Sect.~\ref{subsec:discussion_asymmetry}.\\
\textbf{Hypothesis 1: Direct mass ejection.} The asymmetric structures could correspond to over-densities in the circumbinary disc, generated by gravitational interactions between the companion and the disc itself. Such interactions lead to direct mass transfer through the Lagrangian points L2 and L3. In this scenario, material ejected from the inner binary system would create co-rotating structures with close to Keplerian velocities at the disc location where the elongated clumps are observed.\\
\textbf{Hypothesis 2: Precessing spiral density waves.} Alternatively, the elongated structures similar to spiral-like features, might arise from spiral density waves generated by gravitational interactions between the dwarf and the disc material. These waves form at Lindblad resonances, where the companion's gravitational influence disturbs the disc material, and propagate through the gas and dust \citep{Cuello+2025}. This scenario has been tested numerically and for example it was proposed to explain the symmetric spiral structures observed in the VLT/SPHERE images of the young stellar object \object{HD~100453} \citep{Benisty2017}. Unlike co-rotating structures, spiral density waves would precess at much lower velocities.\\
\textbf{Hypothesis 3: Influence of a third stellar component.} The asymmetries could also result from the gravitational influence of a putative third stellar companion, located several au from the central binary system. This could manifest either as a locally formed gravitationally bound object (e.g. giant planet, brown dwarf) embedded within the disc material at the observed location, or as density enhancements in the circumstellar material caused by gravitational perturbations from a more distant third companion.\\
Long-term IR monitoring of the disc can help discriminate between the three scenarios. Assuming that the large-scale asymmetric structures, especially the SE one, follows a quasi-Keplerian orbit at the estimated distance $R_\mathrm{ref} \sim 10$~au (see Sect.~\ref{subsec:geomodel_asymmetry}), it would imply an orbital period of roughly $11$~years. Comparison of future reconstructed images with new MATISSE observations acquired over a time span of 4--5 years (correspond to $\sim 40\%$ of the SE structure probable orbital period) with the results reported here would clearly indicate whether the structure has moved in a Keplerian orbit (supporting Hyps.~1 and 3) or remained at the same position (strengthen Hyp.~2). If none of these scenarios are favoured by future observations, a fourth one should be considered.\\
\textbf{Hypothesis 4: Radiative transfer effects.} The asymmetries could arise from a combination of radiative transfer effects (e.g. opacity, scattering) and the particular geometry of the circumstellar environment illuminated by the central binary. This last hypothesis would be strengthened if new IR images taken several years later show the same asymmetric structures at identical positions and with similar brightness distributions.

\subsection{Physical and dynamical constraints on $3$~Puppis' disc}\label{subsec:HDanalysis}
Thanks to the MATISSE observations, our study has demonstrated that the asymmetric structures detected in the dusty circumbinary disc of \object{$3$~Pup} are robust across different imaging methods and coherent over three mid-IR spectral channels. The confirmed presence of these asymmetries raises a fundamental astrophysical question: are they driven by local disc instabilities, such as gravitational fragmentation or massive embedded bodies, or do they reflect global tidal perturbations from the central binary? To discriminate between the scenarios proposed in Sect.~\ref{subsec:discussion_origins}, we performed a hydrodynamic (HD) study to assess the physical plausibility of each scenario. While a detailed radiative transfer study would provide more definitive constraints, this preliminary assessment enables us to evaluate competing scenarios and improve our understanding of the physical mechanisms underlying the observed structures.

\subsubsection{Hydrodynamical analysis}\label{subsec:calcul_HDanalysis}
The HD analysis of the SE elongated clump provides critical constraints on the physical mechanisms responsible for the observed asymmetries. Our calculations yield a gas mass estimate of $M_\mathrm{clump} \simeq 3.73\times 10^{-2} \, M_\oplus$ (see Appx.~\ref{app:HD_calcul_massbudget}) and demonstrate extreme gravitational stability at the reference radius $R_\mathrm{ref}$, with Toomre parameter values $Q_\mathrm{ref} \gg 10^4$ (see Appx.~\ref{app:HD_calcul_grav_instability}). These values exceed the classical threshold for axisymmetric gravitational instability ($Q \lesssim 1.5$ according to \citealt{Durisen2007}) by more than four orders of magnitude, effectively ruling out local gravitational collapse processes such as disc clumping due to self-gravity and fragmentation at the inner rim region \citep{Kratter2016}.\\
The mass budget analysis further constrains the nature of the observed structure. Even under optimistic assumptions regarding the gas-to-dust ratio ($\delta = 10^4$, representing a 100-fold increase from our baseline estimate), the clump mass remains insufficient for gravitational binding ($\sim 0.2\,M_\mathrm{Jup}$). This mass falls two to three orders of magnitude below the typical threshold required to form a bound object with a circumplanetary disc \citep{Zhu2015}. This analysis strongly disfavours direct mass ejection leading to gravitational clumping (i.e. Hyp.~1) and makes the presence of embedded giant planets or brown dwarfs at $R_\mathrm{ref}$ highly unlikely (i.e. Hyp.~3). Any such body would require a formation history independent of local gravitational collapse, such as external capture or core accretion from an earlier evolutionary epoch. On the other hand, the viscosity evolution analysis reveals timescales at the clump's position of $9.75 \times 10^4$ yrs that exceed both the central binary orbital period and the estimated period for structures at $R_\mathrm{ref}$ by more than two orders of magnitude. Consequently, any asymmetry observed around the dusty inner rim is unlikely to originate from viscous redistribution of mass, given the long viscous timescales relative to the relevant orbital periods, thereby supporting the persistence of coherent spiral structures.\\
Collectively, these results strongly favour Hyp.~2, meaning that the observed asymmetries are best interpreted as tidally-induced spiral density waves generated by gravitational interactions between the central binary and the circumbinary disc, as described by \citet{ArtymowiczLubow1996} and \citet{Poblete+2019}. This interpretation is consistent with similar structures observed in other binary systems and young stellar objects \citep{Benisty2017}. Detailed calculations characterizing the physical properties of the circumbinary disc, including gravitational stability criteria, viscous evolution timescales, and mass distribution analysis, are presented in Appx.~\ref{app:HD_calcul}. The relevant physical equations and numerical parameters are summarized in Tables~\ref{tab:app_equations} and \ref{tab:app_physparams}.

\subsubsection{Hydrodynamical simulation of the asymmetries}\label{subsec:HD_simu}
\begin{figure}[ht]
    \centering
    \includegraphics[width=0.9\linewidth]{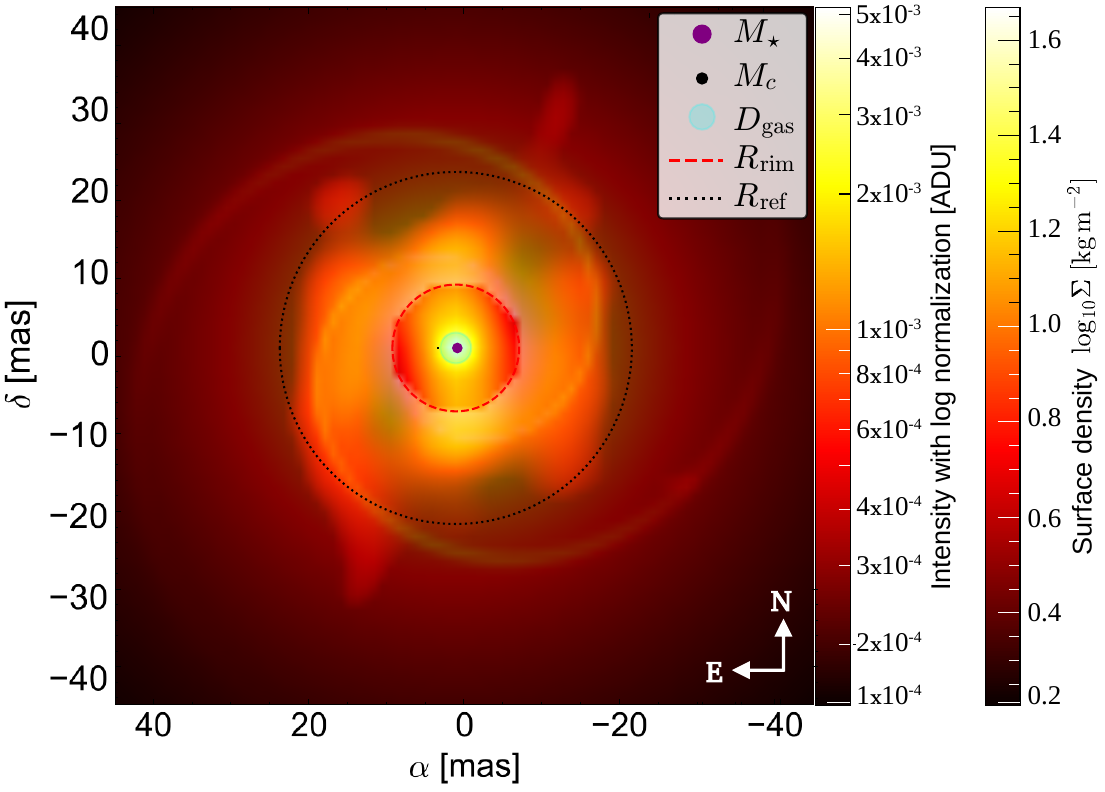}
    \caption{Composite image showing the analytic spiral model superimposed on the reconstructed N-band \texttt{MYTHRA} image of \object{$3$~Puppis}. The logarithmic colour scale represents the observed intensity of the N-band image (normalized to unity and expressed in ADU), while the spiral model is overlaid as a semi-transparent structure representing the local gas surface density normalized to $20\%$ contrast. The red dashed circle marks the inner dust rim $R_\mathrm{rim} \simeq 5.7$~mas, and the black dotted circle indicates the radius of the southeastern asymmetric structure $R_\mathrm{ref} \simeq 16.71$~mas inferred from the L-band \texttt{MYTHRA} image. The curvature and radial extent of the two logarithmic spiral arms closely match the two asymmetries observed in the N-band image, located in the southeastern and northwestern regions of the dusty disc.}
    \label{fig:spiral_obs_model_overlay}
\end{figure}

To assess whether the SE and NW asymmetries detected in the circumbinary disc of \object{$3$~Pup} could originate from a tidally excited spiral density wave, we constructed a two-dimensional analytic model of the disc's surface density field $\Sigma(R,\phi)$ (see formula in Appx.~\ref{app:HD_spiralmodel}). This model assumes a Keplerian disc structure with a background surface density profile following $\Sigma_0(R) \propto R^{-3/2}$, consistent with steady-state viscous evolution \citep{LyndenBell1974}. The reference point is set to the SE clump position in polar coordinates: $R_0 = R_\mathrm{ref} \simeq 16.71$~mas and $\phi_0=\tan^{-1}(y/x)_\mathrm{ref} = -31.74\degr$ (derived from the L-band geometric modelling in Sect.~\ref{subsec:geomodel_asymmetry}). Superimposed on the surface background, we implemented a trailing logarithmic spiral characterized by the pitch angle $\psi$. This logarithmic formulation has been widely adopted in analytical and numerical studies of tidally excited spirals (e.g. \citealp{OgilvieLubow2002, Rafikov2002, Dong2015}), as it naturally captures the characteristic curvature and trailing structure of density waves launched at Lindblad resonances. Drawing from the HD simulations of \cite{Dong2015} and \cite{Zhu2015} who studied tidally perturbed discs around binaries with small mass ratios (i.e. $M_c/M_* \ll 1$) and with an aspect ratio $\epsilon\sim 0.05$, we adopted the following model parameters input values: density contrast amplitude $\Gamma=0.2$ (see justification in Appx.~\ref{app:HD_approx}), Gaussian angular width $\sigma_\phi = 20\degr$ (yielding an effective spiral Gaussian angular thickness of $\displaystyle \widehat{\mathrm{FWHM}}\sim 47\degr$), and $\psi = 15\degr$. To generate similar observational conditions, we projected the vertical optical depth onto the sky plane, assuming axisymmetric thermal emission at a constant temperature (see justifications in Appx.~\ref{app:HD_approx}), to mimic mid-IR radiative transfer. The resulting synthetic surface density distribution of \object{$3$~Pup}'s disc with simulated tidal perturbations is presented in Figure~\ref{fig:app_HDsimu}, while Figure~\ref{fig:spiral_obs_model_overlay} shows the direct overlay of the spiral density model with the N-band reconstructed \texttt{MYTHRA} image (original image plotted at the bottom right panel of Figure~\ref{fig:fit_LMNband}). 
The simulated spirals exhibit a trailing orientation confined within a $\sim15$~au radius, successfully reproducing the elongated clumps in the SE and NW regions of \object{$3$~Pup}'s disc in both N-band \texttt{MYTHRA} and \texttt{SPARCO} images (see Figure~\ref{fig:images}). Importantly, this reproduction requires neither a third companion nor gravitational instability and is fully consistent with the long viscous timescales derived in Sect.~\ref{app:HD_calcul_viscous}. The moderate surface density enhancement (i.e. $\Gamma\sim 20\%$) remains consistent with the dynamical stability conditions derived in Sect.~\ref{app:HD_calcul_grav_instability} (i.e. $Q \gg 1$ ensuring that the disc remains gravitationally stable in the perturbed region). This modelling demonstrates that the observed asymmetries can arise from a tidal response of the circumbinary disc to the inner binary's gravitational potential. The excellent agreement between our analytic spiral morphology and the N-band reconstructed images strongly supports the interpretation of tidally induced spiral waves. This scenario is dynamically stable, hydrodynamically plausible, and consistent with all available observational and theoretical constraints.

\section{Conclusion}\label{sec:concl}
We presented new interferometric observations of the \object{$3$~Pup} system with MATISSE in the L-, M-, and N-bands. The data were analysed for each spectral band using image reconstruction techniques including \texttt{MYTHRA}, a novel statistical framework to produce robust images, and \texttt{SPARCO} algorithm.\\
Our analysis constrained the morphology of the dusty inner rim, revealing that its extent, flattening, and skewness are consistent with previous K-band results from AMBER. Comparison with earlier MIDI observations shows good compatibility, though our MATISSE data provide significantly enhanced information thanks to improved sensitivity, broader spectral coverage, access to longer baselines, and four-telescope configuration enabling CP measurements and hence the detection of asymmetries within the CE. Unlike results from first-generation VLTI instruments, our MATISSE images reveal a far more complex circumstellar morphology around the binary system of \object{$3$~Pup}. While the inner dusty disc structure remains consistent with previous findings, we detect a bright, extended structure towards the SE, located at a projected distance of approximately $10$~au at $631$~pc --roughly three times the dust inner rim radius. A fainter, asymmetric structure is also observed in the opposite direction (i.e. NW), approximately $180\degr$ from the main asymmetric structure revealed in L-band images in the SE region.\\
To explain the nature and origin of these mid-IR asymmetries, we formulated four scenarios: (1) precessing gravity waves in the circumbinary disc due to the inner companion; (2) corotating spirals created by direct mass transfer from the inner system through Lagrangian points; (3) over-density created by an unknown tertiary companion further in the circumbinary disc; (4) radiative transfer effects.\\
Our HD analysis demonstrates extreme gravitational stability across the entire disc ($Q \gg 10^4$) across the disc, ruling out bound massive objects or local collapse due to long viscous timescales ($\sim 10^4$--$10^5$ years) and insufficient gas mass ($\sim 10^{-4} M_\mathrm{Jup}$). Instead, the elongated clumps are best explained by tidally induced spiral density waves from the central binary, supporting the hypothesis (2). The synthetic spiral morphology successfully reproduces the observed structures, strongly supporting this interpretation as dynamically stable and consistent with theoretical HD predictions for circumbinary disc dynamics. Therefore, the referred elongated structures can rather be identified as dusty spiral arcs. \\
Follow-up interferometric observations, as well as a reanalysis of the archival AMBER data and a dedicated radiative transfer modelling, will help discriminate between scenarios. Refining our understanding of this system may shed light on mass-loss processes in evolved binaries and demonstrate the importance of the binary nature of B[e] stars.

\begin{acknowledgements}
Part of this work was supported by the observations collected at the European Organisation for Astronomical Research in the Southern Hemisphere under \emph{ESO} programme(s) 0104.D-0669(A), 0104.D-0015(B), 0104.D-0554(A), 112.25C5.003. MATISSE collaboration is a consortium composed of institutes in France (J.-L. Lagrange Laboratory, Université Côte d'Azur, Observatoire de la Côte d'Azur Observatory, CNRS), Germany (MPIA, MPIfR, and the University of Kiel), the Netherlands (NOVA and the University of Leiden), and Austria (the University of Vienna). MA, AM, ADS, and FM acknowledge the support of the French Agence Nationale de la Recherche (ANR), under grant MASSIF (ANR-21-CE31-0018-01, \url{www.anr-massif.fr}), without which this work would not have been possible. MA also acknowledges the partial supported by the Early-Career Scientific Visitor Programme at ESO Chile (laureate of the 2024 research grant, \url{www.eso.org/sci/activities/santiago/visitors.html}). JV is funded from the Hungarian NKFIH OTKA projects no. K-132406, and K-147380. This work was also supported by the NKFIH NKKP grant ADVANCED 149943. Project no.149943 has been implemented with the support provided by the Ministry of Culture and Innovation of Hungary from the National Research, Development and Innovation Fund, financed under the NKKP ADVANCED funding scheme. JV acknowledges support from the Fizeau exchange visitors programme. The research leading to these results has received funding from the European Union’s Horizon 2020 research and innovation programme under Grant Agreement 101004719 (ORP). PA acknowledges the funding support from the Hungarian NKFIH project No.~K-147380. This research has made use of the Jean-Marie Mariotti Center (JMMC) - MOIO, in particular \texttt{ASPRO2} (\url{www.jmmc.fr/english/tools/proposal-preparation/aspro}), \texttt{OiFitsExplorer}, \texttt{OImaging}, and \texttt{AMHRA} (\url{https://amhra.oca.eu/AMHRA/index.htm}). This work has made use of data from the European Space Agency (ESA) mission \emph{Gaia} (\url{www.cosmos.esa.int/gaia}), processed by the \emph{Gaia} Data Processing and Analysis Consortium (DPAC, \url{www.cosmos.esa.int/web/gaia/dpac/consortium}). Funding for the DPAC has been provided by national institutions, in particular the institutions participating in the \emph{Gaia} Multilateral Agreement. The following services and facilities have also been used: the \texttt{SIMBAD} and \texttt{VIZIER} databases (CDS, Strasbourg, France) and the NASA's Astrophysics data System.
\end{acknowledgements}

\bibliographystyle{aa}
\bibliography{aa56615-25}

\begin{appendix}
\onecolumn
\clearpage
\FloatBarrier
\section{Logbook of $3$~Puppis observations with VLTI/MATISSE}\label{app:logbook}

\begin{table}[ht]
\caption{\label{table:ATs} Standard ATs configurations used in VLTI/MATISSE observations of \object{$3$~Puppis}.} 
\centering
\begin{tabular*}{\textwidth}{@{\extracolsep{\fill}}c c c c c c c c c c}
\hline\hline
\multicolumn{2}{c}{\object{$3$~Pup} Observation} & VLTI & AT stations & \multicolumn{6}{c}{Projected baseline length} \\
\multicolumn{2}{c}{Night | Time (UTC)}& config. & position& \multicolumn{6}{c}{$B$ (m)} \\
\hline
\multicolumn{2}{c}{2020-02-15 | $00^h33^m01^s$} & Small & A0-B2-D0-C1 & $B_\mathrm{min} =$ \textbf{11.26} & 19.89 & 22.54 & 23.31 & 29.36 & 33.80 \\
\multicolumn{2}{c}{2020-02-20 | $04^h54^m13^s$} & Medium & K0-G2-D0-J3 & 36.06 & 56.22 & 56.81 & 57.65 & 79.36 & 92.69\\
\multicolumn{2}{c}{2020-02-27 | $04^h08^m18^s$} & Large & A0-G1-J2-J3 & 51.18 & 83.86 & 103.93 & 113.50 & 118.58 & 130.02\\
\multicolumn{2}{c}{2024-03-10 | $00^h57^m35^s$} & Extended & A0-B5-J2-J6 & 48.47 & 111.96 & 129.11 & 167.34 & 169.42 & $B_\mathrm{max} =$ \textbf{201.43}\\
\multicolumn{2}{c}{2024-03-10 | $04^h46^m33^s$} & Extended & A0-B5-J2-J6 & 48.19 & 73.61 & 91.69 & 144.57 & 167.93 & 184.39\\
\hline
\end{tabular*}
\tablefoot{The starting observation time of \object{$3$~Puppis} is given in the UTC convention and with the units of \textit{hours:minutes:seconds}. $B_\mathrm{max}$ and $B_\mathrm{min}$ are respectively the longest and the shortest projected baseline separation lengths used during the observations. The so called `Large' configuration in this table refers to the old VLTI baseline configuration (config.), that is to say before the ESO call for proposal P112. The `Extended' baseline configuration has been offered to the community since 2023.}
\end{table}

\begin{table}[ht]
\caption{\label{table:obs_MATISSElog} VLTI/MATISSE observing log for \object{$3$~Puppis} performed in low spectral resolution mode (R=30).} 
\centering
\begin{tabular*}{\textwidth}{@{\extracolsep{\fill}}c c c c c c c}
\hline\hline
\multicolumn{2}{c}{\object{$3$~Pup} Observation} & VLTI & ESO Program ID & Seeing & Airmass & Coherence time \\
\multicolumn{2}{c}{Night | Time (UTC)} & config. & & ($''$) & (...) & $\tau_0$ (ms) \\
\hline
\multicolumn{2}{c}{2020-02-15 | $00^h33^m01^s$}& Small & 0104.D-0669(A) & 0.95--0.56 & 1.15--1.15 & 9.5--9.9\\
\multicolumn{2}{c}{2020-02-20 | $04^h54^m13^s$}& Medium & 0104.D-0015(B) & 0.68--0.65 & 1.20--1.20 & 6.3--6.3 \\
\multicolumn{2}{c}{2020-02-27 | $04^h08^m18^s$}& Large & 0104.D-0554(A) & 0.81--0.78 & 1.15--1.15 & 5.8--6.2 \\
\multicolumn{2}{c}{2024-03-10 | $00^h57^m35^s$}& Extended & 112.25C5.003 & 0.64--0.61 & 1.00--1.00 & 6.9--8.0 \\
\multicolumn{2}{c}{2024-03-10 | $04^h46^m33^s$}& Extended & 112.25C5.003 & 0.68--0.68 & 1.50--1.51 & 11.7--11.7 \\
\hline
\end{tabular*}
\tablefoot{The starting observation time of \object{$3$~Puppis} is given in the UTC convention and with the units of \textit{hours:minutes:seconds}. For seeing, airmass, and coherence time columns, the interval of values refer to the measurement of the respective quantity at the start of the observation for the first value and to the end of the observation for the second value.}
\end{table}

\begin{table}[ht]
\caption{\label{table:obs_calib}List of the stars used to calibrate the VLTI/MATISSE observations of \object{$3$~Puppis}.} 
\centering
\begin{tabular*}{\textwidth}{@{\extracolsep{\fill}}c c c c c c c c c c}
\hline\hline
\multicolumn{2}{c}{\object{$3$~Pup} Observation}&\multicolumn{4}{c}{LM calibration star}& \multicolumn{4}{c}{N calibration star} \\
\multicolumn{2}{c}{Night | Time (UTC)}& Name & Type & F$_\mathrm{L}$ (Jy) & $\theta_{\mathrm{UD,L}}$ (mas) & Name & Type & F$_\mathrm{N}$ (Jy) & $\theta_{\mathrm{UD,N}}$ (mas) \\
\hline
\multicolumn{2}{c}{2020-02-15 | $00^h33^m01^s$} & \object{HD 56618} & M2 & 196 & 5.50 & \object{* q Car}& K2 & 46 & 5.23\\
\multicolumn{2}{c}{2020-02-20 | $04^h54^m13^s$} & \object{* 140 Pup} & K3 & 86 & 2.90 & \object{* q Car}& K2 & 46 & 5.23\\
\multicolumn{2}{c}{2020-02-27 | $04^h08^m18^s$} & \object{* eps Ant} & K3 & 88 & 2.86 & \object{* q Car} & K2 & 46 & 5.23\\
\multicolumn{2}{c}{2024-03-10 | $00^h57^m35^s$} & \object{HD 59610} & K3 & 52 & 1.11 & \object{* bet Ori} & B8 & 38 & 2.71\\
\multicolumn{2}{c}{2024-03-10 | $04^h46^m33^s$} & \object{HD 59610}& K3 & 52 & 1.11 &\object{* bet Cnc} & K4 & 35 & 4.91\\
\hline
\end{tabular*}
\tablefoot{Observations have been conducted with the observing sequence CAL-SCI-CAL to better the calibration of the absolute visibility. `CAL' refers to the calibration star and `SCI' for the science target \object{$3$~Puppis}. $\theta_{\mathrm{UD}}$ stands for the measured uniform-disc angular diameter of the calibration stars (the index `L' refers to the measurement made in the L-band and `N' to the N-band). Spectral type, flux and angular diameter of calibrations are provided by the MDFC catalogue \citep{MDFC_Cruzalebes+2019}. The starting observation time is given in the UTC convention and with the units of \textit{hours:minutes:seconds}.}
\end{table}

\begin{table}[ht]
\caption{Interferometric beam characterization for \object{$3$~Puppis} VLTI/MATISSE observations.}
\centering
\begin{tabular}{c c c}
\hline\hline
Spectral& {\small FWHM}$_{x}$& {\small FWHM}$_{y}$\\
band & (mas)& (mas)\\
\hline
L-band & 4.83 & 3.17 \\ 
M-band & 7.20 & 4.75 \\
N-band & 12.80 & 8.45\\
\hline
\end{tabular}
\label{tab:interferometric_beam}
\tablefoot{Results from fitting a two-dimensional (2D) Gaussian distribution to the synthesized beam (a.k.a. the dirty beam) to characterize the angular resolution achieved by the observations in each spectral band of VLTI/MATISSE. The dirty beam is obtained by applying the inverse Fourier transform of the ($u,v$)-coverage. FWHM$_{x}$ and FWHM$_{y}$ represent the full width at half maximum along the $x$- and $y$-axes respectively, for the 2D Gaussian distribution, with a position angle (PA) of $25.23\degr$. This elliptical beam represents the instrument's point spread function (PSF).}
\end{table} 
\FloatBarrier
\clearpage

\FloatBarrier
\section{Image reconstruction results with \texttt{MYTHRA}}\label{app:MYTHRA}
\begin{figure}[ht]
\centering
\includegraphics[width=0.49\linewidth]{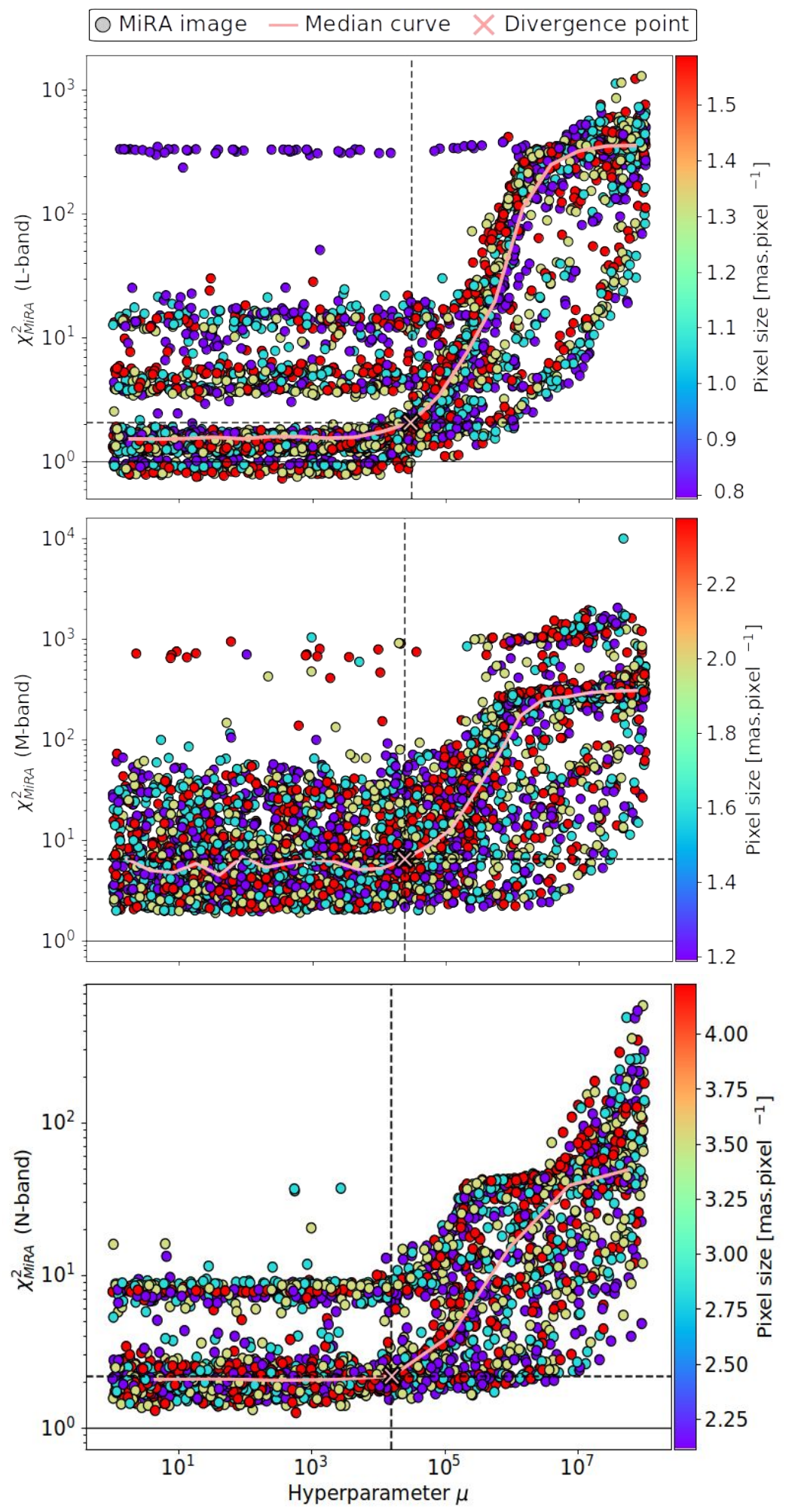}
  \caption{L-curves of the \texttt{MiRA} images reconstructed using the VLTI/MATISSE \object{$3$~Puppis} data, and for each spectral band. The median curve resulting from the data binning performed by \texttt{MYTHRA} is shown by the solid black line, and the pink cross indicates the inflection point, corresponding to the optimal hyperparameter value. Each coloured dot represents one of the 3360 reconstructed images generated by the \texttt{PYRA} grid. The colourmap reflects the pixel size value used to reconstruct the circumstellar environment. \textit{Top plot:} L-band. \textit{Center plot:} M-band. \textit{Bottom plot:} N-band.}
  \label{fig:Lcurve}
\end{figure}
\FloatBarrier
\clearpage
\onecolumn

\FloatBarrier
\begin{table}[ht]
\caption{Field of view and angular resolution characteristics probed by VLTI/MATISSE for the observation of $3$~Puppis.}\centering
\begin{tabular*}{\textwidth}{@{\extracolsep{\fill}}c c c c c c c}
\hline\hline
Spectral& Coverage & Centred wavelength & Angular resolution & Super resolution & FoV$_\text{AT}$ & FoV$_\text{interf}$ \\
band & $\Delta\lambda$ ($\mu$m) & $\lambda_0$ ($\mu$m)& $\theta_\text{r}$ (mas) & $\theta^+$ (mas) & (mas) &(mas)\\
\hline
\noalign{\smallskip}
L-band & $3.0 - 3.9$ & $3.45$ & $1.77$ & $0.88$ & $593$ & $63$ \\ 
M-band & $4.5 - 4.9$ & $4.75$ & $2.43$ & $1.22$ & $816$ & $87$ \\
N-band & $8.0 - 12$ & $10.0$ & $5.12$ & $2.56$ & $2\,291$ & $183$\\
\hline
\end{tabular*}
\label{tab:resolution_band}
\tablefoot{The field of view (FoV) coherently perceived by VLTI/MATISSE is referred to as the interferometric FoV, noted FoV$_\text{interf}$. In contrast, the photometric FoV, noted FoV$_\text{AT}$, corresponds to the field of view that is limited by the aperture $D$ of a single telescope dish (e.g. in the case of VLTI/MATISSE, this aperture refers to the size of an Auxiliary Telescope dish) and is modulated by a coefficient $c$ due to the pinhole size, following the relation FoV$_\text{AT} = c\,\lambda_0 / D$ \citep{MATISSE_Lopez+2022}.}
\end{table} 

\begin{table}[ht]
\caption{Statistic figures obtained with \texttt{MYTHRA} applied on the resulting \texttt{PYRA} grid of 3360 images reconstructed from \object{$3$~Puppis} data.}
\begin{tabular*}{\textwidth}{@{\extracolsep{\fill}}c c c c c c c c c}
\hline\hline 
\noalign{\smallskip}
Spectral& L-curve inflection & $\chi^2_\mathrm{MiRA}$ & Selected & Images kept & \multicolumn{4}{c}{Averaged image characteristics}\\
band & point ($\mu^+|\,\chi_+^2) $ & criterion & subset &for the mean&Pixel size $(\text{mas}\cdot\text{pixel}^{-1})$&$\chi^2_\mathrm{global}$&$\chi^2_\mathrm{vis}$&$\chi^2_\mathrm{clos}$\\
\hline
\noalign{\smallskip}
L-band & $2.95 \cdot 10^4\,|\, 2.1$ & $[0.5 - 3.0]$ & 224 & 181 &0.397& 7.6 & 8.4 & 6.4\\
M-band & $2.41 \cdot10^4\,|\, 6.5$ & $[0.5 - 10]$ & 226 & 80 &0.594&5.3 & 6.8 & 2.9\\
N-band & $1.57 \cdot 10^4\,|\, 2.2$ & $[0.5 - 5.0]$ & 253 & 253 &1.056& 2.2 & 3.3 & 0.6\\
\hline
\end{tabular*}
\tablefoot{The $\chi^2_\mathrm{MiRA}$ interval corresponds to the reduced chi-squared value returned by \texttt{MiRA} in output for the reconstructed image. Whereas $\chi^2_\mathrm{global}$ corresponds to the weighted sum of the $\chi^2_\mathrm{vis}$ from the squared visibility fit and of the $\chi^2_\mathrm{clos}$ from the closure phase fit both resulting from the comparison between the simulated data computed from the averaged images and the VLTI/MATISSE data.}
\label{tab:outputs_MYTHRA}
\end{table}
\FloatBarrier

\section{Comparison between VLTI/MIDI data and VLTI/MATISSE N-band data}\label{app:MIDI_MATISSE_comp}
\FloatBarrier
\begin{table}[ht]
\caption{\label{table:Nband_geometricmodel}Fitted parameters from the chromatic geometric model applied to VLTI/MIDI and VLTI/MATISSE N-band observations of \object{$3$~Puppis}.}
\centering
\begin{tabular*}{\textwidth}{@{\extracolsep{\fill}}c |c c c c c c | c}
\hline \hline
\noalign{\smallskip}
VLTI& \multicolumn{6}{c|}{EG comp.} & PS comp.\\
instrument&\multicolumn{2}{c}{{\small FWHM} (mas)} & \multicolumn{2}{c}{$e$ (...)}& PA ($\degr$)& $f_\mathrm{EG}$ ($\%$ total flux) & $f_\mathrm{PS}$ ($\%$ total flux)\\
& $8.0\,\mu$m & $12\,\mu$m & $8.0\,\mu$m & $12\,\mu$m & achromatic & achromatic & achromatic \\
\hline
\noalign{\smallskip}
MIDI &$16 \pm 1$ & $28\pm 2$ & $1.00 \pm 0.05$ &$1.70 \pm 0.30$ & $22 \pm 59$ & $77 \pm 3$& $23 \pm 3$\\
\hline
\noalign{\smallskip}
MATISSE & $23.0\pm 0.5$& $30.0\pm 0.5$ & $1.50 \pm 0.05$ & $1.20 \pm 0.05$ & $-14 \pm 1$& $83 \pm 1$&$17 \pm 1$ \\
\hline
\end{tabular*}
\tablefoot{The chromatic geometric modelling consists of an centred elliptical Gaussian disc component (EG comp.) and a point source component (PS comp.) which describes the fully resolved component at the centre of the image. Two chromatic parameters set to free are the full width at half maximum size of the major-axis of the Gaussian disc (FWHM) and its elongation ratio ($e$). The disc major-axis' position angle (PA) and the relative flux contribution of the disc ($f_\mathrm{EG}$) are the two remaining free parameters of the model fitting that were set as achromatic. The relative flux for the point source component is inferred from the formula $f_\mathrm{PS} = 100\% - f_\mathrm{EG}$.}
\end{table}

\clearpage
\twocolumn
\begin{figure}[h]
\centering
\includegraphics[width=0.9\linewidth]{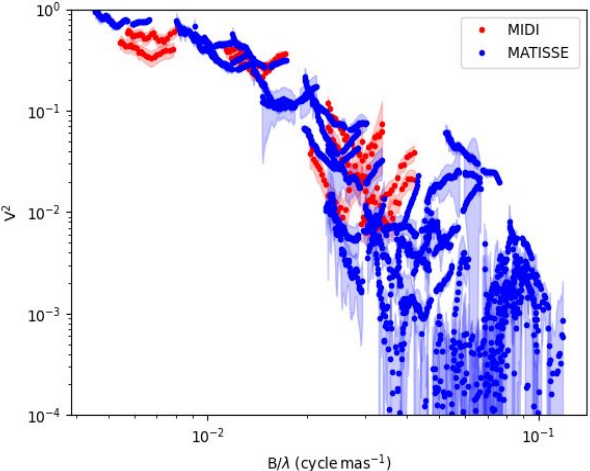}
\caption{Comparison in the N-band of the VLTI/MATISSE data (in blue) to the VLTI/MIDI data (in red) of \object{$3$~Puppis} by plotting their squared visibilities data (V$^2$) in logarithmic scale as a function of spatial frequency.}
\label{fig:MIDIandMATISSEdata}
\end{figure}
\FloatBarrier

\begin{figure}[h]
\centering
\includegraphics[width=1\linewidth]{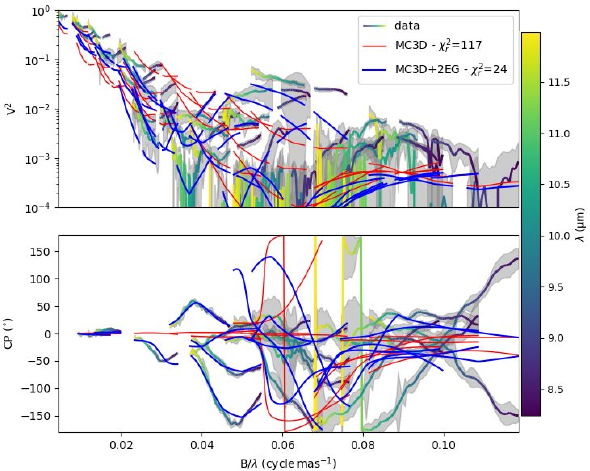}
\caption{Comparison in the N-band between the VLTI/MATISSE data of \object{$3$~Puppis} (in colour gradient from green to yellow), the best-fit \texttt{MC3D} model from \cite{Meilland+2010} (in red) composed of a single centred Gaussian component plus an extended background, and the same \texttt{MC3D} model combined with two elliptical Gaussian components (i.e. \{MC3D + 2EG\} model, in blue) to account for the bright southeastern asymmetry and the fainter northwestern one. The latter model provides a significantly better fit to the MATISSE interferometric observables than the original \texttt{MC3D} model constrained by MIDI data, with a reduced chi-squared of 24 versus 117. \textit{Top plot:} Squared visibilities (V$^2$) in logarithmic scale as a function of spatial frequency. \textit{Bottom plot:} Closure phases (CPs) as a function of spatial frequency.}
\label{fig:MATISSE_M3CDmodel}
\end{figure}

\section{Hydrodynamic framework}\label{app:HD}
\subsection{Approximations adopted}\label{app:HD_approx}

\subsubsection{Mean molecular weight value ($\mu$)}\label{app:HD_approx_molecularweight} 
We adopt a mean molecular weight of $\mu = 2.3$, corresponding to the standard value for molecular gas of solar mass composition (i.e. about $70\%$ of hydrogen, $28\%$ of helium, and $2\%$ of heavier elements). This value is widely used in the literature for protoplanetary and circumbinary disc models where the gas is predominantly molecular and weakly ionized \citep{Hayashi1981, DAlessio1998, Draine2011}. The choice on this value is justified by the typical thermochemical conditions in the outer regions of irradiated discs (i.e. temperatures $T \lesssim 3000$~K, densities $n \gtrsim 10^8\,\mathrm{cm^{-3}}$, and a very low ionization fraction). Under these conditions, hydrogen is primarily in molecular form (H$_2$), helium remains neutral (He I), and metals are mostly bound in grains or remain singly ionized, yielding a stable effective $\mu \simeq 2.3$. Departures from this value may occur only in highly irradiated, ionized, or dissociative environments.

\subsubsection{Disc locally isothermal}\label{app:HD_approx_isothermal}
Local isothermality holds when the cooling timescale $t_\mathrm{cool}$ is much shorter than the dynamical (orbital) timescale $t_\mathrm{dyn} \simeq \Omega^{-1}$. In dusty, irradiated discs such as \object{$3$~Pup}, the thermal inertia of the gas is low and radiative cooling is efficient due to tight coupling with dust grains. Consequently, temperature fluctuations are rapidly equilibrated with the local irradiation field, leading to a vertically isothermal structure at each radius \citep{RudenPollack1991}.

\subsubsection{Geometrically thin-disc}\label{app:HD_approx_thindisc}
The geometrically thin-disc approximation holds when the vertical scale height is much smaller than the radial distance, hence an aspect ratio of $\epsilon \ll 1$. As shown in Appx.~\ref{app:HD_calcul_thindiscmodel}, we obtain an aspect ratios of $\epsilon \simeq 0.044$ at the inner rim ($R_\mathrm{rim} \sim 3.6$~au at $631$~pc) and $\epsilon \simeq 0.023$ at the SE clump position ($R_\mathrm{ref} \sim 10$~au at $631$~pc). These values are fully consistent with flared, passively irradiated discs in hydrostatic equilibrium and justify the use of the thin-disc approximation \citep{KenyonHartmann1987, ChiangGoldreich1997}.

\subsubsection{Gaussian vertical density profile ($\rho$)}\label{app:HD_approx_densityprofile} 
We assume the disc to be locally isothermal and geometrically thin, in line with standard models of circumbinary and protoplanetary discs \citep{ChiangGoldreich1997, Dullemond2001}. Under these assumptions, the condition of vertical hydrostatic equilibrium leads to a Gaussian vertical distribution of the gas density:
\begin{equation*}
    \rho(z) = \rho_0 \times \exp\left( -\frac{z^2}{2\,H^2} \right)
\end{equation*}
where $\rho_0$ is the midplane density at a given radius (e.g. at the inner rim), $z$ is the vertical coordinate above the midplane, and $H$ is the (pressure) scale height as defined in Eq.~\ref{tab:app_equations}\textcolor{blue}{.5}. This formulation assumes that the temperature is vertically constant and only varies radially (i.e. $T = T(R)$), consistent with the locally isothermal approximation. This simplification greatly facilitates both thermodynamical and hydrodynamical modelling and remains valid in the inner $\sim10$--$20$~au of circumbinary discs around evolved B[e] stars, where stellar irradiation dominates over viscous heating.

\subsubsection{Radial thickness ($w$)}\label{app:HD_approx_thickness} 
We define the radial thickness of the SE clump as the local vertical scale height of the disc, under the geometrically thin-disc approximation, yielding $w \equiv H_\mathrm{ref}$. Using the values computed in Appx.~\ref{app:HD_calcul_thindiscmodel}, we obtain then $w \simeq 0.23$~au, which defines the natural vertical and radial confinement scale of spiral density waves in hydrostatic equilibrium. This choice avoids overestimating the mass by restricting the integration region to physically meaningful disc layers.

\subsubsection{Angular width ($\Delta\phi$)}\label{app:HD_approx_angularwidth}  
The angular width defines the azimuthal extent of the asymmetric structure identified in the SE quadrant of the disc, referred to as the SE elongated clump (see Sect.~\ref{subsec:discussion_asymmetry}), in the region of $R_\mathrm{ref}$. This quantity represents the angular opening of the overdensity along its curved geometry as reconstructed in the MATISSE images (see Figure~\ref{fig:images}). To measure it, we analyse the unconvolved M-band \texttt{MYTHRA} reconstructed image (middle left panel of Figure~\ref{fig:fit_LMNband}) by transforming from Cartesian to polar coordinates. In particular, we focus on pixels within the radial interval $[R_\mathrm{min},R_\mathrm{max}] = [13, 19]$~mas, corresponding to the region where the SE elongated clump extends in the M-band image. Then, for each radius $r$ within these boundaries, we convert pixel intensities to angular positions $\phi$ along circles centred on the image centre and with a radius equal to the given $r$. After collecting all angular pixel values and applying a $3\sigma$ threshold to identify robustly clump elements, we finally determine the angular width by subtracting the minimum angular position from the maximum one. The calculation reveals that the asymmetry spans a well-localized arc corresponding to a continuous spiral segment subtending approximately $\Delta\phi \simeq 34\degr$ in azimuth. \\
This width is consistent with typical angular spans of spiral arms observed in circumbinary discs with moderate density contrast (e.g. $\Delta\phi\in[20,40]\degr$ according to \citealp{Dong2015}). Therefore, we adopt $\Delta\phi\sim30\degr$ ($= \pi/6$ rad) as a representative estimate for the extent of the overdense region.

\subsubsection{Clump area ($A_\mathrm{clump}$)}\label{app:HD_approx_area}  
The total area occupied by the SE clump is computed as a sector of disc annulus at radius $R_\mathrm{ref}$, with radial thickness $w$, and angular width $\Delta\phi$, yielding the formulation given in Eq.~\ref{tab:app_equations}\textcolor{blue}{.9}. This definition provides the projected surface area of the clump and allows the estimation of its mass content $M_\mathrm{clump}$ as a first-order approximation (see Eq.~\ref{tab:app_equations}\textcolor{blue}{.10}).

\subsubsection{Density contrast amplitude ($\Gamma$)}\label{app:HD_approx_contrast}  
The density contrast amplitude is a dimensionless parameter that quantifies the maximum enhancement of the surface density of a spiral arm (e.g. the SE elongated clump) with respect to the local background surface density profile of the disc. Observationally, focusing for instance on the SE elongated clump, the contrast represents the relative intensity of the clump $I_\mathrm{clump}$ above the intensity level set by the disc background $I_\mathrm{bkd}$ and can be expressed as $\displaystyle \Gamma = (I_\mathrm{clump} - I_\mathrm{bkd})/I_\mathrm{bkd}$. From our analysis of the M-band \texttt{MYTHRA} image, during the angular width $\Delta\phi$ measurement, we estimated the $\Gamma\sim 20\%$ for the SE clump. In the analytical framework, $\Gamma$ appears as a model parameter in the spiral density wave equation $\Sigma(R,\phi)$ (see Appx.~\ref{app:HD_spiralmodel}).

\subsection{Hydrodynamic calculations}\label{app:HD_calcul}
\subsubsection{Gravitational instability within the disc}\label{app:HD_calcul_grav_instability}
To compute the gravitational stability based on the Toomre theoretical framework \citep{Toomre1964}, whose formula is given in Eq.~\ref{tab:app_equations}\textcolor{blue}{.4}, we first need to determine the surface density $\Sigma_\mathrm{rim}$ occupied by the gas at the inner rim, assuming axisymmetric geometry. Using Eq.~\ref{tab:app_equations}\textcolor{blue}{.1}, we obtain a value of $\Sigma_\mathrm{rim} \simeq 38.2\ \mathrm{kg}\cdot \mathrm{m^{-2}}$. We can then infer the angular frequency at the inner rim using Eq.~\ref{tab:app_equations}\textcolor{blue}{.3}, given that the disc material follows Keplerian motion, yielding $\Omega_\mathrm{rim} \simeq 9.0 \times 10^{-8}$~ rad$\cdot$s$^{-1}$. Finally, the Toomre parameter at the dusty inner rim reads $Q_\mathrm{rim} \simeq 2.4 \times 10^4$.\\
Similarly, we want to assess the gravitational stability at larger radii, specifically at $R_\mathrm{ref}$. In the context of viscously evolving discs, we assume a power-law surface density profile defined as $\Sigma(R) \propto R^{-3/2}$ \citep{LyndenBell1974}. Hence, the density at $R_\mathrm{ref} = 10$~au (at $631$~pc) can be inferred as follows:
\begin{equation*}
\Sigma_\mathrm{ref} \simeq \Sigma_\mathrm{rim} \times \left( \frac{R_\mathrm{ref}}{R_\mathrm{rim}} \right)^{-3/2} \simeq 8.25\,\mathrm{kg}\cdot\mathrm{m^{-2}}.
\end{equation*}
The corresponding angular frequency gives $\Omega_\mathrm{ref} \simeq 6.15 \times 10^{-8}\, \mathrm{rad}\cdot\mathrm{s^{-1}}$, yielding to a Toomre parameter equal to $Q_\mathrm{ref} \simeq 7.5 \times 10^5$.

\subsubsection{Verification of the thin-disc model for $3$~Pup}\label{app:HD_calcul_thindiscmodel}
Under the locally isothermal approximation (see justifications in Appx.~\ref{app:HD_approx}) and using the quantities estimated in Appx.~\ref{app:HD_calcul_grav_instability}, we can infer the vertical scale height of the dusty disc at the inner rim through Eq.~\ref{tab:app_equations}\textcolor{blue}{.5}, yielding $H_\mathrm{rim} \simeq 0.157$~au. From Eq.~\ref{tab:app_equations}\textcolor{blue}{.6}, the corresponding aspect ratio at the inner rim is $\epsilon_\mathrm{rim} \simeq 0.044$, which is consistent with values expected for passive irradiated discs in hydrostatic equilibrium, typically ranging between 0.03 and 0.1 \citep{ChiangGoldreich1997, Dullemond2001}. This aspect ratio confirms that the disc geometry supports the thin-disc model approximation adopted in deriving the Toomre parameter in Appx.~\ref{app:HD_calcul_grav_instability}. To verify whether this approximation remains valid at larger radii, we extend our analysis to $R_\mathrm{ref}$.\\
Following the same approach, we first compute the angular frequency at $R_\mathrm{ref}$ using Eq.~\ref{tab:app_equations}\textcolor{blue}{.3}, obtaining $\Omega_\mathrm{ref}=6.15 \times 10^{-8}\,\mathrm{rad}\cdot\mathrm{s^{-1}}$. This yields a vertical scale height of $H_\mathrm{ref} \simeq 0.23$~au and an aspect ratio of $\epsilon_\mathrm{ref} \simeq 0.023$. These values demonstrate that while the disc exhibits modest flaring with increasing radius, it remains geometrically thin, from the inner rim, out to $R_\mathrm{ref}\sim10$~au, thereby validating our thin-disc approximation throughout this region.

\subsubsection{Viscosity evolution and timescale}\label{app:HD_calcul_viscous}
The disc’s kinematic viscosity is parametrized using the \citet{ShakuraSunyaev1973} $\alpha$-prescription as defined in Eq.~\ref{tab:app_equations}\textcolor{blue}{.7}. Adopting a standard viscosity parameter $\alpha = 10^{-2}$ in the case of moderately turbulent protoplanetary or post-AGB discs \citep{Hartmann1998, Rafikov2016}, we obtain a viscosity of $\nu_\mathrm{rim} \simeq 4.98 \times 10^{11}\, \mathrm{m^2}\cdot\mathrm{s^{-1}}$ at the inner rim, corresponding to a viscous timescale of $t_\mathrm{visc,rim} \simeq 1.85 \times 10^4\, \mathrm{yrs}$ using Eq.~\ref{tab:app_equations}\textcolor{blue}{.8}. Extending this analysis to $R_\mathrm{ref}$ using the same viscosity law, we find respectively $\nu_\mathrm{ref} \simeq 7.28 \times 10^{11}\,\mathrm{m^2}\cdot\mathrm{s^{-1}}$ and $t_\mathrm{visc,ref} \simeq 9.75 \times 10^4\, \mathrm{yrs}$.\\
These viscous timescales exceed by more than two orders of magnitude both the orbital period of the central binary ($P_\mathrm{orb} \simeq 137.52$~days, see Table~\ref{tab:params_prop_lPup_litterature}) and the period inferred for a putative a third body orbiting at $R_\mathrm{ref}$ ($11$~yrs, see Sect.~\ref{subsec:discussion_origins}).

\subsubsection{Mass budget of the SE asymmetry}\label{app:HD_calcul_massbudget}
To assess whether the observed SE elongated clump could originate from a bound object, we need to estimate the total gas mass contained within its area (i.e. in $R_\mathrm{ref}$ region). We model the clump as extending over a sector of angular width $\Delta \phi = 30^\circ$ and radial thickness $w \sim H_\mathrm{ref}$ (see justifications in Appx.~\ref{app:HD_approx}). Using Eq.~\ref{tab:app_equations}\textcolor{blue}{.9}, the defined geometry for the SE clump yields an area of $A_\mathrm{clump} \simeq 2.70 \times 10^{22}\, \mathrm{m^2}$. Combined with the local gas surface density $\Sigma_\mathrm{ref} \simeq 8.25\,\mathrm{kg\cdot m^{-2}}$ (see Appx.~\ref{app:HD_calcul_grav_instability}), we obtain a total gas mass enclosed within the clump sector of $M_\mathrm{clump} \simeq 3.73\times 10^{-2}\, M_\oplus \simeq 1.18 \times 10^{-4}\,M_\mathrm{Jup}$ based on Eq.~\ref{tab:app_equations}\textcolor{blue}{.10} (see justifications in Appx.~\ref{app:HD_approx}). 

\subsection{Hydrodynamic simulation of a spiral density model}\label{app:HD_spiralmodel}
\begin{figure}[h]
    \centering
    \includegraphics[width=1\linewidth]{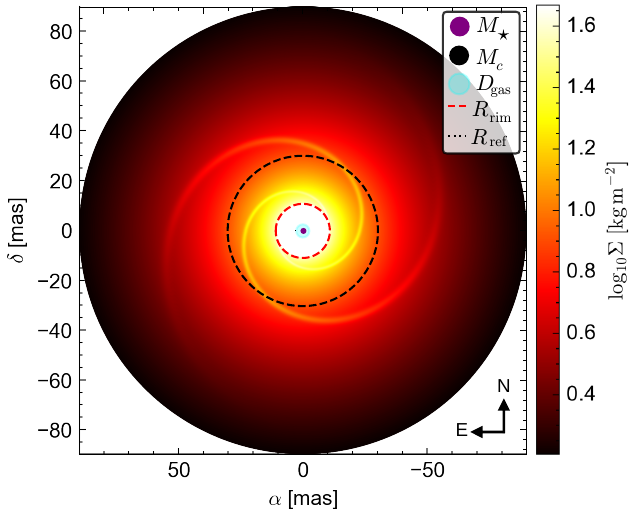}
    \includegraphics[width=1\linewidth]{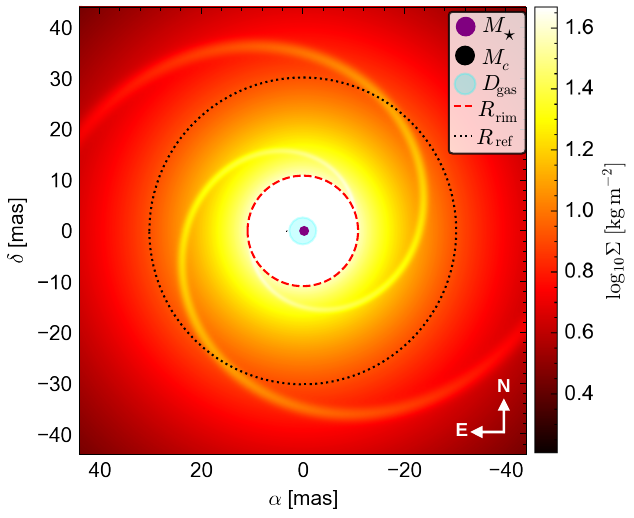}
    \caption{Synthetic surface density map $\Sigma$ derived from an analytic spiral density model simulating tidally-induced perturbations in the circumbinary disc of \object{$3$~Puppis}. The model incorporates two trailing spiral arm perturbations caused by the binary companion. The red dashed circle marks the inner dust rim $R_\mathrm{rim} \simeq 5.7$~mas ($\sim 3.6$~au at $631$~pc), and the black dotted circle indicates the radius of the southeastern asymmetric structure $R_\mathrm{ref} \simeq 16.71$~mas ($\sim 10$~au at $631$~pc) inferred from the L-band \texttt{MYTHRA} image. Binary star positions are marked with dot symbols (purple for the supergiant star and black for the low-mass companion) scaled to the correct mass ratio and projected separation (stellar radii not to scale). The gas disc extension $D_\mathrm{gas} \sim 4.39$~mas ($\sim 0.88$~au at $631$~pc) is displayed as a blue circle to scale. Both panels use logarithmic colour scaling to represent the local gas surface density normalized to $20\%$ contrast. \textit{Left plot:} Wide field of view ($160 ~\mathrm{mas} \times 160~\mathrm{mas}$) in sky-plane Cartesian coordinates $(\alpha,\delta)$. \textit{Right plot:} Same simulation with a field of view matching the VLTI/MATISSE M-band \texttt{MYTHRA} image ($87~\mathrm{mas} \times 87~\mathrm{mas}$).}
    \label{fig:app_HDsimu}
\end{figure}

The spiral density wave is defined analytically by a two-dimension model corresponding to the disc's surface density field $\Sigma(R,\phi)$, in polar coordinates $(R,\phi)$, as follows:
\begin{eqnarray*}
    \Sigma(R,\phi)&=&\Sigma_0(R) \times \left[\,1+ \Gamma \times \exp \left(-\frac{[\phi-\Phi(R)]^2}{2\, \sigma_\phi^2}\right)\,\right]\\
    \mathrm{with} \quad \Phi(R)&=&\phi_0 + \frac{1}{\tan \psi} \times \ln \left(\frac{R}{R_0}\right)
\end{eqnarray*}
where $\psi$ is the pitch angle that defines the logarithmic spiral parametrisation, $\Sigma_0(R)$ represents the disc surface density profile (i.e. without the intensity contribution of spiral structure), $\Gamma$ is the density contrast amplitude of the spiral wave (see definition in Appx.~\ref{app:HD_approx}), $\sigma_\phi$ is the Gaussian azimuthal envelope angular width, $R_0$ is the reference radius, and $\phi_0$ is the initial phase of the spiral density wave. The resulting synthetic surface density model with spiral-induced perturbations is shown in Figure~\ref{fig:app_HDsimu}.

\onecolumn
\subsection{Dynamical parameters and physical conditions}\label{app:HD_tab}
\begin{table}[ht]
\caption{\label{tab:app_equations}Dynamical quantities evaluated at the inner rim ($R_\mathrm{rim}$) and at the distance where the southeastern asymmetry is ($R_\mathrm{ref}$) in \object{$3$~Puppis}' disc.}
\begin{tabular*}{\textwidth}{@{\extracolsep{\fill}}c l l c c c}
\hline\hline
\noalign{\smallskip}
Eq. n$\degr$ & Quantity& Symbol (Unit)& Expression & Value at $R_\mathrm{rim}$ & Value at $R_\mathrm{ref}$ \\
\noalign{\smallskip}
\hline
\noalign{\smallskip}
0& Distance from the centre of mass& $R$ (au)& $\displaystyle \frac{d_\mathrm{GeDR2}\,(\mathrm{pc}) \times R\,(\mathrm{mas})}{10^{3}}$ & $3.6$ & $10$ \\
\noalign{\smallskip}
\noalign{\smallskip}
1&Surface density& $\Sigma$ (kg$\cdot$m$^{-2}$)& $\displaystyle \frac{M_\mathrm{gas}}{2\pi\,R^2}$ & $38.2$ & $8.25$\\
\noalign{\smallskip}
\noalign{\smallskip}
2&Sound speed&$c_s$ (m$\cdot$s$^{-1}$)& $\displaystyle \sqrt{\frac{k_B\,T_\mathrm{rim}}{\mu\,m_H}}$ & $2\,116$& $2\,116$ \\
\noalign{\smallskip}
\noalign{\smallskip}
3&Keplerian angular frequency& $\Omega$ (rad$\cdot$s$^{-1}$) & $\displaystyle \sqrt{\frac{G\,M_\mathrm{tot}}{R^3}}$ & $9.0 \times 10^{-8}$ & $6.15 \times 10^{-8}$\\
\noalign{\smallskip}
\noalign{\smallskip}
4&Toomre parameter& $Q$ (...) & $\displaystyle \frac{c_s\,\Omega}{\pi\,G\,\Sigma}$ & $2.4 \times 10^4$ & $7.5 \times 10^5$ \\
\noalign{\smallskip}
\noalign{\smallskip}
5&Vertical scale height& $H$ (au)& $\displaystyle \frac{c_s}{\Omega}$ & $0.157$& $0.23$ \\
\noalign{\smallskip}
\noalign{\smallskip}
6&Aspect ratio& $\epsilon$ (...)& $\displaystyle \frac{H}{R}$ & $0.044$ & $0.023$ \\
\noalign{\smallskip}
\noalign{\smallskip}
\noalign{\smallskip}
7&Kinematic viscosity& $\nu$ (m$^2\cdot$s$^{-1}$)& $\alpha \times c_s \times H$ & $4.98 \times 10^{11}$& $7.28 \times 10^{11}$ \\
\noalign{\smallskip}
\noalign{\smallskip}
8&Viscous timescale& $t_\mathrm{visc}$ (yrs)& $\displaystyle \frac{R^2}{\nu}$ & $1.85 \times 10^4$ & $9.75 \times 10^4$ \\
\noalign{\smallskip}
\noalign{\smallskip}
9&Clump area& $A_\mathrm{clump}$ (m$^2$)& $\displaystyle \Delta\phi \times R_\mathrm{ref} \times w$ & -- & $2.70 \times 10^{22}$ \\
\noalign{\smallskip}
\noalign{\smallskip}
\noalign{\smallskip}
10&Gas mass within a clump& $M_\mathrm{clump}$ ($M_\oplus$)& $\Sigma_\mathrm{ref} \times A_\mathrm{clump}$ & -- & $3.73 \times 10^{-2}$ \\
& & $M_\mathrm{clump}$ ($M_\mathrm{Jup}$)& & -- & $1.18 \times 10^{-4}$ \\
\noalign{\smallskip}
\hline
\end{tabular*}
\label{tab:derived}
\end{table}

\begin{table*}[ht]
\centering
\caption{\label{tab:app_physparams}Physical parameters and assumptions adopted for the dynamical analysis of the circumbinary disc around \object{$3$~Puppis}.}
\begin{tabular*}{\textwidth}{@{\extracolsep{\fill}}l l l c}
\hline\hline
\noalign{\smallskip}
Quantity & Symbol (Unit) & Value & References \\
\noalign{\smallskip}
\hline
\noalign{\smallskip}
Gas to dust mass ratio& $\delta$ (...) & $100$ & \citet{Kluska2018, Bujarrabal2013} \\
\noalign{\smallskip}
Dust mass &$M_\mathrm{dust}$ ($M_\sun$)& $3.5 \times 10^{-7}$ & from Table~\ref{tab:params_prop_lPup_litterature} \\
\noalign{\smallskip}
Gas mass &$M_\mathrm{gas}$ ($M_\sun$)& $3.5 \times 10^{-5}$ & $= \delta \cdot M_\mathrm{dust}$ \\
\noalign{\smallskip}
Dusty inner rim radius &$R_\mathrm{rim}$ (au)& $3.6$ & from Table~\ref{tab:params_prop_lPup_litterature} \\
\noalign{\smallskip}
Temperature at the dusty inner rim& $T_\mathrm{rim}$ (K) & $1\,250$ & from Table~\ref{tab:params_prop_lPup_litterature} \\
\noalign{\smallskip}
Mean molecular weight& $\mu$ (...)& $2.3$ & \citet{Hayashi1981}, \citet{DAlessio1998} \\
\noalign{\smallskip}
Mass of the \object{$3$~Puppis} system & $M_\mathrm{tot}$ ($M_\sun$)& $9.55$ & $=M_* + M_c$, derived from Table~\ref{tab:params_prop_lPup_litterature} \\
\noalign{\smallskip}
Sound speed &$c_s$ (m$\cdot$s$^{-1}$)& $2\,116$ & derived from Table~\ref{tab:app_equations} Equation~2\\
\noalign{\smallskip}
Kinematic viscosity parameter& $\alpha$ (...)& $10^{-2}$ & \citet{Hartmann1998, Rafikov2016} \\
\noalign{\smallskip}
Clump radial thickness& $w$ (au)& $0.23$ & $=H(R_\mathrm{ref})$, derived from Table~\ref{tab:app_equations} Equation~5\\
\noalign{\smallskip}
Clump angular width sector & $\Delta\phi$ ($\degr$)& $30$ & derived from M-band \texttt{MYTHRA} image\\
\noalign{\smallskip}
Density contrast amplitude & $\Gamma$ ($\%$)& $20$ & derived from M-band \texttt{MYTHRA} image\\
\noalign{\smallskip}
Spiral pitch angle & $\psi$ ($\degr$)& $15$& \citet{OgilvieLubow2002, Rafikov2002, Dong2015}\\
\noalign{\smallskip}
Gaussian angular width &$\sigma_\phi$ ($\degr$)& 20 & \citet{Dong2015,Zhu2015} \\
\hline
\end{tabular*}
\label{tab:assumptions}
\end{table*}
\end{appendix}
\end{document}